\documentclass[letterpaper,journal]{IEEEtran}
\usepackage{amsmath,amsfonts}
\usepackage[lined,ruled,commentsnumbered]{algorithm2e}
\usepackage{bm} 
\usepackage{amssymb}
\usepackage{array}
\usepackage[caption=false,font=footnotesize,labelfont=rm,textfont=rm,subrefformat=parens,labelformat=parens]{subfig}

\usepackage{textcomp}
\usepackage{stfloats}
\usepackage{url}
\usepackage{verbatim}
\usepackage{graphicx}
\usepackage{cite}
\usepackage{multirow}
\usepackage{colortbl}
\usepackage[colorlinks=true, linkcolor=blue, citecolor=blue,urlcolor=black]{hyperref}
\usepackage{booktabs}
\usepackage[table,xcdraw]{xcolor}
\usepackage{makecell}
\newcolumntype{M}[1]{>{\centering\arraybackslash}m{#1}}

\begin{document}

\title{Age of Information Minimization in UAV-Enabled Integrated Sensing and Communication Systems}

\author{Yu Bai,~\IEEEmembership{Graduate Student Member,~IEEE,}  
        Yifan Zhang,~\IEEEmembership{Graduate Student Member,~IEEE,} \\
        Boxuan Xie,~\IEEEmembership{Graduate Student Member,~IEEE,} 
        Zheng Chang,~\IEEEmembership{Senior Member, IEEE,} \\
        Yanru Zhang,~\IEEEmembership{Senior Member,~IEEE,} 
        Riku J\"antti,~\IEEEmembership{Senior Member, IEEE,}
        and Zhu~Han,~\IEEEmembership{Fellow,~IEEE.}

\thanks{Y. Bai, Y. Zhang, B. Xie, and R. J\"antti are with the Department of Information and Communications Engineering, Aalto University, 02150 Espoo, Finland (e-mail: yu.bai@aalto.fi, yifan.1.zhang@aalto.fi, boxuan.xie@aalto.fi, riku.jantti@aalto.fi)}

\thanks{Z. Chang is with the School of Computer Science and Engineering, University of Electronic Science and Technology of China, Chengdu 611731, China, and also with the Faculty of Information Technology, University of Jyv\"askyl\"a, 40014 Jyv\"askyl\"a, Finland (e-mail: zheng.chang@jyu.fi)
}

\thanks{Y. Zhang is with the School of Computer Science and Engineering,
 University of Electronic Science and Technology of China, Chengdu 611731,
 China, and also with the Shenzhen Institute for Advanced Study, University
 of Electronic Science and Technology of China, Shenzhen 611731, China
 (e-mail:yanruzhang@uestc.edu.cn).}

\thanks{Z. Han is with the Department of Electrical and Computer Engineering, University of Houston, Houston, TX 77004 USA, and also with the Department of Computer Science and Engineering, Kyung Hee University, Seoul 446-701, South Korea (email: hanzhu22@gmail.com)}
}

\markboth{Journal of \LaTeX\ Class Files,~Vol.~14, No.~8, August~2021}%
{Shell \MakeLowercase{\textit{et al.}}: A Sample Article Using IEEEtran.cls for IEEE Journals}


\maketitle

\begin{abstract}
Unmanned aerial vehicles (UAVs) equipped with integrated sensing and communication (ISAC) capabilities are envisioned to play a pivotal role in future wireless networks due to their enhanced flexibility and efficiency. However, jointly optimizing UAV trajectory planning, multi-user communication, and target sensing under stringent resource constraints and time-critical conditions remains a significant challenge. To address this, we propose an Age of Information (AoI)-centric UAV-ISAC system that simultaneously performs target sensing and serves multiple ground users, emphasizing information freshness as the core performance metric. We formulate a long-term average AoI minimization problem that jointly optimizes the UAV's flight trajectory and beamforming. To tackle the high-dimensional, non-convexity of this problem, we develop a deep reinforcement learning (DRL)-based algorithm capable of providing real-time decisions on UAV movement and beamforming for both radar sensing and multi-user communication. Specifically, a Kalman filter is employed for accurate target state prediction, regularized zero-forcing is utilized to mitigate inter-user interference, and the Soft Actor-Critic algorithm is applied for training the DRL agent on continuous actions. The proposed framework adaptively balances the trade-offs between sensing accuracy and communication quality. Extensive simulation results demonstrate that our proposed method consistently achieves lower average AoI compared to baseline approaches.
\end{abstract}


\section{Introduction}
\label{sec:Introduction}

    \IEEEPARstart{U}{nmanned} aerial vehicles (UAVs) have become an essential component in the evolution towards the sixth-generation (6G) and future wireless networks, due to their flexible deployment, high mobility, and independence from terrestrial infrastructure \cite{geraci2022will}. These characteristics make UAVs highly suitable for diverse and challenging missions, such as disaster relief, precision agriculture, border surveillance, and temporary wireless coverage restoration  \cite{wang2021disaster, radoglou2020compilation,berrahal2016border,bai2025dynamic}. Increasingly, UAVs are expected to simultaneously perform wireless communication and sensing tasks within these scenarios \cite{wu2021comprehensive}. Traditionally, these functionalities have relied on separate hardware platforms and dedicated frequency resources, which pose significant constraints in terms of payload, energy efficiency, and overall system complexity on UAV platforms \cite{wilson2021embedded}.

    To overcome these limitations, Integrated Sensing and Communication (ISAC) has emerged as a promising paradigm wherein radar sensing and wireless communication functionalities share hardware resources and frequency bands \cite{liu2022integrated}. Equipping UAVs with ISAC capabilities enables concurrent execution of sensing and communication tasks, thereby providing a compact, efficient solution particularly suited for resource-constrained platforms \cite{mu2023uav}. However, the simultaneous integration of sensing and communication introduces a fundamental challenge: effectively balancing and optimizing these two competing functionalities. The ability to quantitatively evaluate how well sensing and communication tasks are jointly executed becomes essential for optimizing overall system performance, particularly given the dynamic and mobile nature of UAV deployments \cite{lu2024integrated}.

    Typical performance assessments for UAV-enabled ISAC systems have focused on the traditional communication or sensing performance \cite{meng2023uav}. However, they fall short in capturing the distinct demands of time-critical scenarios common in UAV missions such as disaster response and intelligent transportation systems, where the timely delivery of fresh information is crucial. In such contexts, the value of information rapidly deteriorates over time, making timeliness a critical performance dimension. Addressing this issue, Age of Information (AoI), a metric gaining traction in time-sensitive applications such as multi-access edge computing (MEC), has emerged as a particularly appropriate measure \cite{yates2021age, chiariotti2021peak}. Motivated by this advantage, adopting AoI as a primary performance metric in UAV-based ISAC systems allows for a more comprehensive and precise assessment of timely information delivery in dynamic and time-sensitive missions.


    \subsection{Related Work}

    Existing research on UAV-enabled ISAC systems typically focuses on optimizing either communication-centric or sensing-centric performance metrics. Communication-centric studies generally aim to enhance communication performance under sensing constraints, such as maximizing communication users' sum-rate, average-rate, or minimum-rate \cite{meng2022throughput, lyu2022joint, liu2024resource, zhang2024joint, deng2023beamforming, deng12024integrated}, guaranteeing quality of service (QoS) \cite{khalili2024efficient}, or maximizing the achievable rates under the security constraints \cite{deng2024joint}. Conversely, sensing-centric works focus on improving target sensing efficiency, optimizing metrics like the successful tracking ratio \cite{zhou2022integrated}, the Cramér-Rao bound (CRB) \cite{jiang2024uav, jing2024isac}, the radar probing error \cite{liu2024radar}, and the radar estimation rate \cite{liu2024uav2, liu2025integrated}. Additionally, UAV-specific metrics, including collision avoidance and energy efficiency, have been explored to enhance real-world applicability \cite{chen2025reinforcement, khalili2024efficient}. Among various optimization strategies, joint optimization of UAV trajectory and beamforming has been widely recognized as essential for fully utilizing UAV spatial degrees of freedom and achieving superior system performance \cite{meng2022throughput, lyu2022joint, deng12024integrated, deng2024joint, zhang2024joint, deng2023beamforming}.    

    However, the communication or sensing-oriented metrics employed in existing studies typically neglect the timeliness requirements inherent to UAV missions, particularly in dynamic, time-critical scenarios. To better capture the temporal dimension in UAV-based missions, the AoI metric has recently emerged as a suitable measure of information freshness in time-sensitive applications such as MEC and disaster response \cite{yates2021age, chiariotti2021peak}. Several recent works have adopted AoI to assess UAV-ISAC system performance \cite{zhu2023aoi,zhou2024optimizing, mei2025aoi, 10891341}. For instance, Zhu et al. \cite{zhu2023aoi} reduced AoI through UAV trajectory optimization while neglecting wireless resource allocation. Zhou et al. \cite{zhou2024optimizing} extended AoI optimization to multi-UAV scenarios, yet maintained temporal alternation between sensing and communication without spatial beamforming. Mei et al. \cite{mei2025aoi} considered AoI under interference constraints, yet their work remained limited to single-link scenarios. Liu et al. \cite{10891341} introduced onboard computation resources but ignored multi-user spatial multiplexing capabilities. Although these studies initially explored the benefits of AoI, they did not fully exploit the joint trajectory and spatial beamforming optimization potential, thereby leaving significant space for improvement.

    To effectively address these joint optimization challenges, selecting an appropriate method is critical. UAV-enabled ISAC optimization problems often exhibit strong variable coupling, non-convexity, and multi-objective characteristics, thus limiting the flexibility and scalability of classical techniques such as alternating optimization, heuristic methods, and convex approximation \cite{meng2022throughput, zhang2024joint, deng2023beamforming}. Recently, deep reinforcement learning (DRL) has emerged as a promising alternative, enabling end-to-end policy learning and adaptive decision-making in dynamic environments without explicit problem decomposition \cite{bai2023toward}. DRL has successfully optimized UAV trajectory and user tracking \cite{chen2025reinforcement, liu2025integrated}, jointly addressed user association, trajectory planning, and power allocation \cite{10086052}, and integrated clustering-based user assignment with trajectory and beamforming optimization \cite{10663708}. However, existing DRL approaches predominantly focus on conventional metrics, leaving AoI-centric optimization in UAV-enabled ISAC underexplored.

    \subsection{Motivation and Contribution}

    Despite advances in UAV-enabled ISAC systems, several research gaps remain that motivate our work.
    
    First, the potential of AoI as a metric for time-critical UAV-enabled ISAC tasks remains insufficiently explored. Most prior works optimize communication performance under sensing constraints \cite{meng2022throughput, lyu2022joint} or sensing accuracy under data rate guarantees \cite{zhou2022integrated, jiang2024uav}. Meanwhile, existing UAV-based ISAC systems that adopt AoI as a performance metric are still limited to data collection scenarios and have not considered time-critical applications \cite{zhu2023aoi,zhou2024optimizing, mei2025aoi, 10891341}. 

    Second, the beamforming capabilities of UAV-mounted antenna arrays remain unexplored in existing AoI-driven UAV-ISAC systems. Recent AoI-centric UAV studies mainly focus on reducing latency via trajectory planning or time-division protocols alone \cite{zhu2023aoi, zhou2024optimizing, mei2025aoi, 10891341}, yet the spatial degrees of freedom offered by multi-antenna ISAC platforms have remained unexplored. Without joint beamforming and trajectory design, these approaches cannot fully exploit the antenna array’s ability to serve multiple ground users and the sensing target simultaneously. 

    Third, existing optimization methods lack flexibility to adapt to different user distributions and dynamic target motions under AoI-driven UAV-ISAC settings. Alternating optimization needs to fix one set of variables while optimizing another, limited to solving different user distributions or target dynamics change \cite{meng2022throughput, zhang2024joint, deng2023beamforming}. Although some DRL-based solutions have emerged for UAV path planning and rate maximization, they typically overlook AoI-related objectives and may not scale well to scenarios with multiple users or rapidly moving targets \cite{chen2025reinforcement, liu2025integrated}. These gaps underscore the need for a more adaptive framework capable of handling the interdependence of UAV trajectory and beam control in real-time. 

    To bridge these research gaps, this paper proposes a UAV-enabled ISAC framework that exploits DRL to jointly optimize the UAV’s trajectory, multi-user beamforming, and radar sensing, with the explicit objective of minimizing AoI in dynamic settings. Unlike conventional ISAC research that focuses primarily on throughput or detection accuracy, we emphasize information freshness to reflect the real-time requirements of disaster relief, transportation safety, and other time-critical scenarios. The key contributions are summarized below:

    \begin{itemize}
        \item \textbf{AoI‐Centric UAV‐ISAC System for Time‐Critical Missions.} We introduce a UAV-enabled ISAC system explicitly optimized for minimizing AoI, defined as the age of the most recent target state updates successfully received by ground users. AoI captures the freshness of information, crucial for time-critical missions like disaster relief. Meanwhile, AoI in our system is influenced simultaneously by the allocation of sensing and communication resources, inherently reflecting a critical balance between these two components. The proposed system comprehensively integrates UAV dynamics, active sensing, and multi-user communication tasks, thereby ensuring efficient, timely, and balanced resource utilization.

        \item \textbf{Spatially-Aware Beamforming for Multi-User Communication and Target Sensing.} We introduce a joint sensing-communication waveform design leveraging the spatial degrees of freedom provided by the UAV-mounted uniform planar array (UPA) antenna within the AoI-centric UAV-ISAC system. This spatial resource allocation strategy enables precise beamforming towards multiple users and mobile targets simultaneously, significantly improving AoI performance. Unlike existing AoI-centric UAV-ISAC approaches, our spatially-aware waveform design effectively balances and enhances both communication reliability and sensing accuracy.

        \item \textbf{DRL-Based Joint UAV Trajectory and Beamforming.} We propose a novel DRL-based algorithm to jointly optimize the UAV’s trajectory and beamforming within the UAV-enabled ISAC system. By formulating the problem as a finite Markov Decision Process (MDP), our DRL agent learns optimal UAV displacement and priority-based beam allocation policies. A subsequent post-processing stage ensures feasible power distributions and beam directions, employing regularized zero-forcing to mitigate multiuser interference and utilizing a Kalman filter for accurate tracking of the dynamic target. We employ the Soft Actor-Critic (SAC) method to efficiently train our DRL agent for continuous action spaces, ensuring stable and effective learning performance.

    \end{itemize}

    
    The remainder of this paper is organized as follows. Section \ref{sec:System_Model} presents the system model and problem formulation of the proposed AoI‐centric UAV‐ISAC system. Section \ref{sec:Proposed_Solution} details the proposed DRL-based approach. Section \ref{sec:SimulationResults} provides a comprehensive evaluation of the proposed system and algorithm. Finally, we conclude the paper in Section \ref{sec:Conclusion}.

\section{System Model and Problem Formulation}
\label{sec:System_Model}

\begin{figure}
        \centering
        \includegraphics[width=0.9\columnwidth]{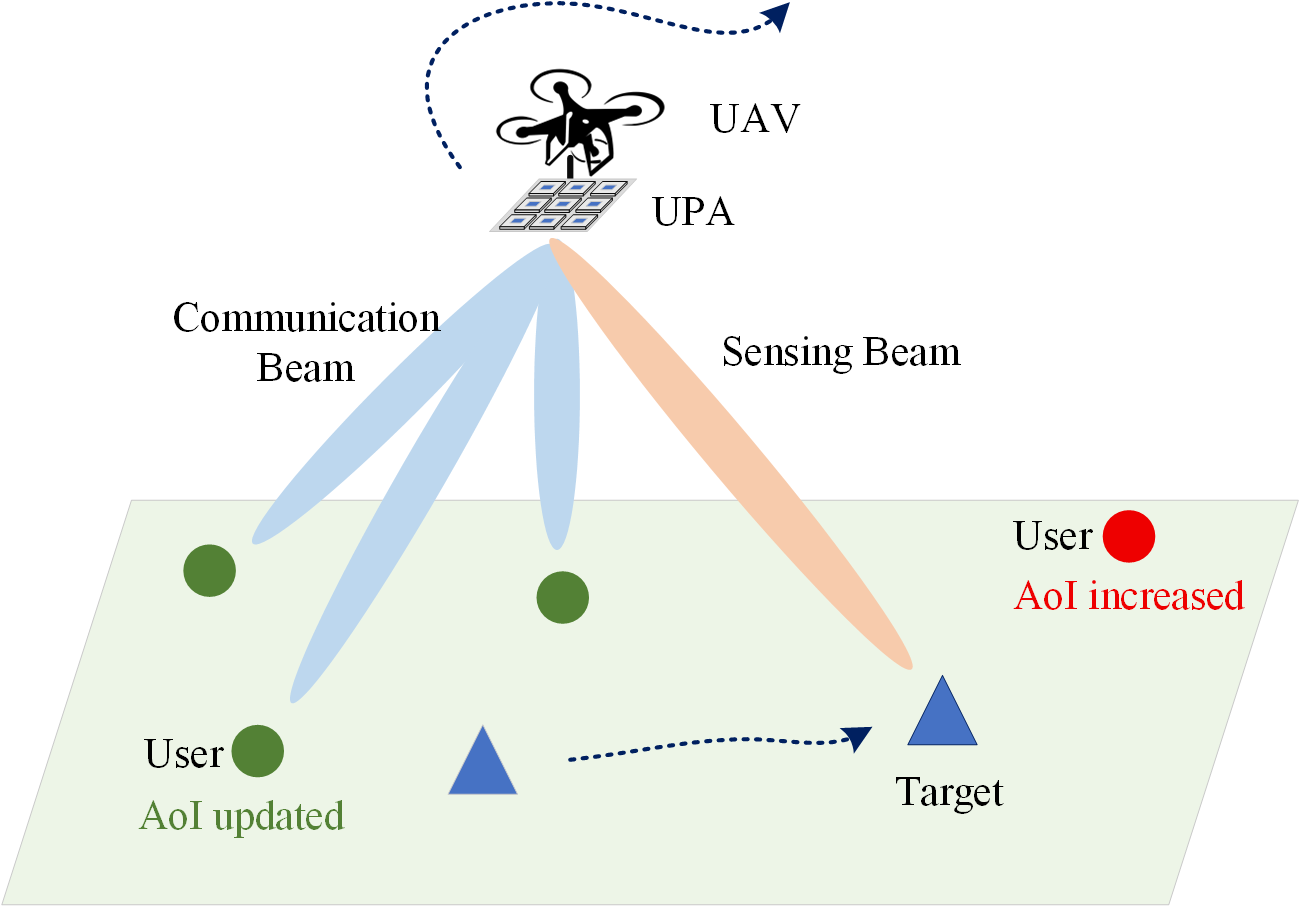}
        \caption{Illustration of the proposed UAV-enabled ISAC system.}
        \label{fig:system}
    \end{figure}

    We consider a UAV-enabled ISAC system, where a single UAV is deployed to simultaneously perform moving target sensing and downlink communication for $K$ ground users, as illustrated in Fig.~\ref{fig:system}. The set of ground users is denoted by $\mathcal{K} = \{1,2, \dots, K\}$. The location of user $k$ is denoted by $\mathbf{p}_k = [x_k, y_k, 0]$, which is assumed to be known to the UAV in advance. The target is a mobile entity that moves within the same horizontal plane. Its position at time slot $n$ is denoted as $\mathbf{p}_{\mathcal{T}}[n] = [x_{\mathcal{T}}[n], y_{\mathcal{T}}[n], 0]$, which is unknown and time-varying, and must be estimated by the UAV through active sensing. We adopt a discrete-time model consisting of $N$ time slots, each lasting a duration $\delta_t$, and index each slot by $n \in \{1,2, \dots, N\}$.

    \subsection{UAV and Antenna Array Geometry}
    The UAV is equipped with a UPA antenna consisting of $M = M_x \!\times\! M_y$ isotropic elements, spaced at half a wavelength in both axes, i.e.\ $d_x = d_y = \lambda/2$. Let $f_c$ denote the carrier frequency and $c$ the speed of light, such that $\lambda = c/f_c$. Each array element has a gain of $G_{\mathrm{elem}}$. The position of the UAV at time slot $n$ is denoted as $\mathbf{p}_u[n] = [x_u[n], y_u[n], H]$, with the fixed altitude $H$. The UAV is assumed to move with a maximum horizontal velocity of $v_{\max}$.
    
    The angle-of-departure (AoD) from the UAV to user~$k$ is characterized by the azimuth $\psi_k[n]\in[-\pi,\pi]$ and elevation $\vartheta_k[n]\in[0,\pi/2]$, calculated as
    \begin{equation}
        \begin{aligned}
        \psi_k[n] &= \operatorname{atan2} \left(y_k - y_u[n], x_k - x_u[n]\right), \\
        \vartheta_k[n] &= \arccos \left(\frac{H}{\|\mathbf{p}_u[n] - \mathbf{p}_k\|}\right).
        \end{aligned}
        \label{eq:angle}
    \end{equation}

   Based on the AoD pair, the corresponding beam steering vector to user $k$ at time slot $n$ is
    \begin{equation}
    \begin{aligned}
        \mathbf{a}(\psi_k[n],\vartheta_k[n]) = \mathbf{a}_y(\psi_k[n],\vartheta_k[n]) \otimes \mathbf{a}_x(\psi_k[n],\vartheta_k[n]),
    \end{aligned}
    \label{eq:steer_total}
    \end{equation}
    where $\otimes$ represents the Kronecker product of the steering vectors along the x-axis and y-axis
    \begin{equation}
    \begin{aligned}
        \mathbf{a}_x(\psi_k[n],\vartheta_k[n]) =
            \big[
            & 1, e^{-j \frac{2\pi d_x}{\lambda} \sin(\vartheta_k[n]) \cos(\psi_k[n])}, \dots, \\
            & e^{-j \frac{2\pi d_x}{\lambda} (M_x - 1) \sin(\vartheta_k[n]) \cos(\psi_k[n])}
            \big],
    \end{aligned}
    \end{equation}
    \begin{equation}
    \begin{aligned}
        \mathbf{a}_y(\psi_k[n],\vartheta_k[n]) =
        \big[
        & 1, e^{-j \frac{2\pi d_y}{\lambda} \sin(\vartheta_k[n]) \sin(\psi_k[n])}, \dots, \\
        & e^{-j \frac{2\pi d_y}{\lambda} (M_y - 1) \sin(\vartheta_k[n]) \sin(\psi_k[n])}
        \big].
    \end{aligned}
    \end{equation}
    
\subsection{Communication Model}
\label{sec:CommModel}
    Inspired by \cite{liu2018mu,sturm2011waveform,li2016optimum}, we adopt a superimposed waveform design to enable simultaneous multi-user downlink communication and target sensing. Specifically, at time slot $n$, the total transmit signal from the UAV is given by
    \begin{equation}
    	\mathbf{x}[n]
    	\;=\;
    	\underbrace{\sum_{k=1}^{K}\mathbf{w}_k[n]\;u_k[n]}_{\text{multi-user communication}}
    	\;+\;
    	\underbrace{\mathbf{w}_{\mathcal{T}}[n]\;u_{\mathcal{T}}[n]}_{\text{sensing/probing}},
    	\label{eq:superimposed_tx}
    \end{equation}
    where $u_k[n]$ and $u_{\mathcal{T}}[n]$ are complex baseband waveforms (e.g., orthogonal code sequences or spread-spectrum signals). The beamforming vectors $\mathbf{w}_k[n] \in \mathbb{C}^{M \times 1}$ and $\mathbf{w}_{\mathcal{T}}[n] \in \mathbb{C}^{M \times 1}$ are defined as
    \begin{equation}
        \mathbf{w}_i[n] = \sqrt{P_i[n]}\,\mathbf{v}_i[n],
        \label{eq:beam}
    \end{equation}
    where $i \in \{\mathcal{T}, 1, \ldots, K\}$, and $\mathbf{v}_i[n]$ is the unit-norm beamforming direction vector, i.e., $\|\mathbf{v}_i[n]\|_2 = 1$.
    
    Since all beams share the same RF front-end, the total instantaneous transmit power at time slot $n$ must satisfy
    \begin{equation}
        \sum_{k}P_k[n]+P_{\mathcal T}[n]\le P_{\max}.
    \end{equation}

    Moreover, we assume that any Doppler shift due to UAV mobility is perfectly compensated \cite{meng2022throughput, xing2009motion}. The wireless channel from the UAV to user $k$ follows a free-space path loss model described by the Friis transmission equation:
    \begin{equation}
        \beta_k[n] = \frac{G_{\mathrm{elem}}\, G_{\mathrm{user}}\, \lambda^2}{(4\pi\, d_{u,k}[n])^2},
        \label{eq:pathloss}
    \end{equation}
    where $d_{u,k}[n] = \|\mathbf{p}_u[n] - \mathbf{p}_k\|$ is the Euclidean distance between the UAV and user~$k$ at time slot~$n$. Here, $G_{\mathrm{elem}}$ and $G_{\mathrm{user}}$ denote the antenna gains of the UAV and the single-antenna user~$k$, respectively. The corresponding channel vector from the UAV to user~$k$ is expressed as
    \begin{equation}
        \mathbf h_k^{\mathrm H}[n]=\sqrt{\beta_k[n]}\,e^{-j\tfrac{2\pi}{\lambda}d_{u,k}[n]}\, \mathbf a^{\mathrm H}\bigl(\psi_k[n],\vartheta_k[n]\bigr).
    \end{equation}
    Under the assumption that the sensing waveform is orthogonal to the despreading code of user $k$ \cite{10664619}, it does not contribute to the interference at the receiver. Hence, the signal-to-interference-plus-noise ratio (SINR) at user~$k$ is given by
    \begin{equation}
    \Gamma_k[n]
    =\frac{|\mathbf h_k^{\mathrm H}[n]\mathbf w_k[n]|^{2}}
           {\sum_{k'\neq k}
              |\mathbf h_k^{\mathrm H}[n]\mathbf w_{k'}[n]|^{2}
            +\xi_k^2},
    \label{eq:SINR}
    \end{equation}
    where $\xi_k^2$ is the receiver noise power at user~$k$. A threshold $\Gamma_{\mathrm{th}}$ is established to guarantee reliable communication.

    \subsection{Sensing Model}
    \label{sec:SensingModel}

    To detect and track the moving target, the UAV employs a dedicated probing beam characterized by the steering vector $\mathbf{v}_{\mathcal{T}}[n]$. The directional gain towards the target is quantified by the one-way array factor gain:
    \begin{equation}
        G_{\mathrm{AF}}[n] = \left|\mathbf{a}^{\mathrm{H}}(\psi_{\mathcal{T}}[n],\vartheta_{\mathcal{T}}[n]) , \mathbf{v}_{\mathcal{T}}[n]\right|^{2}, 
    \end{equation}
    where $(\psi_{\mathcal{T}}[n], \vartheta_{\mathcal{T}}[n])$ denote the azimuth and elevation angles from the UAV to the target, computed as described in~\eqref{eq:angle}. The steering vector $\mathbf{a}^{\mathrm{H}}(\psi_{\mathcal{T}}[n], \vartheta_{\mathcal{T}}[n])$ is derived similarly to~\eqref{eq:steer_total}.

    To model the received signal power, we extend the conventional single-antenna radar equation~\cite{levanon2004radar} by incorporating the element gain $G_{\mathrm{elem}}$ and the squared array factor (applied for both transmission and reception), yielding:
    \begin{equation}
    P_{r}[n]
    =\frac{P_{\mathcal T}[n]\,
            \bigl(G_{\mathrm{elem}}\,G_{\mathrm{AF}}[n]\bigr)^{2}
            \,\lambda^{2}\,\sigma_{\mathcal T}}
           {(4\pi)^{3}\,
            \|\mathbf p_u[n]-\mathbf p_{\mathcal T}[n]\|^{4}},
    \label{eq:RadarEq}
    \end{equation}
    where $\sigma_{\mathcal{T}}$ denotes the radar cross-section (RCS) of the target.

    Assuming coherent pulse integration at the receiver, the matched filter operates over a bandwidth $B$ and integrates $N_p$ pulses \cite{levanon2004radar}. The thermal noise power is given by $k_{\mathrm{B}} T_0 B F$, where $k_{\mathrm{B}}$ is the Boltzmann’s constant, $T_0$ is the noise temperature, and $F$ is the receiver noise figure. Consequently, the post-detection signal-to-noise ratio (SNR) per coherent processing interval is

    \begin{equation}
    \mathrm{SNR}_{p}[n]
    =\frac{P_{r}[n]}{k_{\mathrm B}T_{0}B\,F}\;N_p.
    \label{eq:SNRp}
    \end{equation}
    
    \subsubsection{Measurement model and reliability test}
    The radar processor converts the range–bearing measurements to horizontal Cartesian coordinates, yielding
    \begin{equation}
    \mathbf z[n]
    =\mathbf p_{\mathcal T}^{\mathrm h}[n]\;+\;\boldsymbol\Xi[n],
    \quad
    \boldsymbol\Xi[n]
    \sim \mathcal{N}\bigl(\mathbf 0,\mathbf R[n]\bigr).
    \label{eq:meas_z}
    \end{equation}
    where $\mathbf p_{\mathcal T}^{\mathrm h}[n]=(x_{\mathcal T},y_{\mathcal T})$ is the target horizontal coordinates. \(\boldsymbol{\Xi}[n]\) denotes Gaussian measurement noise with zero mean and covariance matrix \(\mathbf{R}[n]\). 
    
    Following the Cramér–Rao lower bound (CRLB) analysis in~\cite{jiang2024uav,jing2024isac}, the measurement covariance matrix can be approximated by
    \begin{equation}
    \mathbf R[n]
    =\frac{\sigma_{0}^{2}}
           {\mathrm{SNR}_{p}[n]+\varepsilon}
     \,\mathbf{I}_2,
    \quad
    \sigma_{0}=\frac{c}{\sqrt{8}\,\pi B},
    \label{eq:Rcov}
    \end{equation}
    where $\sigma_{0}$ sets the high-SNR bound on range accuracy and $B$ is the radar signal bandwidth. \(\mathbf{I}_2\) is the \(2\times 2\) identity matrix. $\mathrm{SNR}_{p}[n]$ is the per-pulse signal–to–noise ratio derived in~\eqref{eq:SNRp} and a tiny constant $\varepsilon>0$ avoids division by zero in numerical computations. 

    Let \(\sigma_{\mathrm{req}}\) denote the position accuracy threshold, defined as the maximum allowable 1-$\sigma$ horizontal position error  (e.g., \(1\,\mathrm{m}\)). A Cartesian measurement is considered reliable when the standard deviation of each coordinate given by the diagonal of its covariance matrix \(\mathbf R[n]\) satisfies 
    \begin{equation}
        \sqrt{\operatorname{diag}\!\bigl(\mathbf R[n]\bigr)}
        \;\le\;\sigma_{\mathrm{req}} .
    \end{equation}
    where $\operatorname{diag}$ denotes the element-wise variance of the 2D measurement noise. This is equivalent to the SNR threshold based on~\eqref{eq:Rcov}
    \begin{equation}
        \mathrm{SNR}_{p}[n]\;\ge\;\mathrm{SNR}_{\mathrm{th}} \triangleq
        \bigl(\tfrac{\sigma_{0}}{\sigma_{\mathrm{req}}}\bigr)^{2}.
        \label{eq:SNR_threshold}
    \end{equation}
    
    \subsubsection{KF-based Target State Estimation}
    We stack the target’s horizontal kinematics in 
    \begin{equation}
          \mathbf{s}_{\mathcal{T}}[n]=
      \bigl[x_{\mathcal T}[n],\,y_{\mathcal T}[n],\,v_x[n],\,v_y[n]\bigr]^{\mathsf T}.
    \end{equation}
    where \((x_{\mathcal{T}},\,y_{\mathcal{T}})\) and \((v_x,\,v_y)\) denote the target's position and velocity, respectively.  Assuming a nearly-constant-velocity (CV) model with time slot length \(\delta_t\), we have 
    \begin{equation}
      \mathbf{s}_{\mathcal{T}}[n+1]=\mathbf{F}\,\mathbf{s}_{\mathcal{T}}[n]+\mathbf{w}[n],\;
      \mathbf{w}[n]\!\sim\!\mathcal N(\mathbf 0,\mathbf{Q}),
        \label{eq:cv_model_short}
    \end{equation}
    where
    \begin{equation}
        \mathbf{F}
        =
        \begin{bmatrix}
            1 & 0 & \delta_t & 0\\
            0 & 1 & 0 & \delta_t\\
            0 & 0 & 1 & 0\\
            0 & 0 & 0 & 1
        \end{bmatrix},
        \quad
        \mathbf{Q}=q_{0}^{2}\mathbf I_{4},
        \label{eq:F_Q_definition}
    \end{equation}
    where \(q_{0}^{2}\) tunes the model-mismatch level, and \(\mathbf{I}_4\) is the \(4\times 4\) identity matrix.    
    The measurement in \eqref{eq:meas_z} can be written as
    \begin{equation}
        \mathbf{z}[n]
        \;=\;
        \mathbf{H}\,\mathbf{s}_{\mathcal{T}}[n]
        \;+\;
        \boldsymbol{\Xi}[n],
        \quad
        \mathbf{H}
        =
        \begin{bmatrix}
           1 & 0 & 0 & 0\\
           0 & 1 & 0 & 0
        \end{bmatrix}.
    \end{equation}

    To predict the target’s state (position and velocity), we employ a standard discrete-time Kalman Filter (KF)~\cite{bar1990tracking, blackman1999design}. Let
    \(
      \hat{\mathbf{s}}_{\mathcal{T}}[n]
      =
      [\,\hat{x}_{\mathcal{T}}[n],\,\hat{y}_{\mathcal{T}}[n],\,\hat{v}_{x}
      [n],\,\hat{v}_{y}[n]\,]^\mathsf{T}
    \)
    be the state estimate at slot \(n\), and \(\mathbf C[n]\) its error covariance. Each slot involves:
    
    \begin{itemize}
      \item \emph{Prediction step}:
      \begin{align}
         \hat{\mathbf{s}}_{\mathcal{T}}^{-}[n] & =\mathbf F\hat{\mathbf{s}}_{\mathcal{T}}^{-}[n-1], \\
        \mathbf C^{-}[n]&=\mathbf F\mathbf C[n-1]\mathbf F^{\mathsf T}+\mathbf Q.
      \end{align}
      \item \emph{Gate on SNR}:
        If \(\mathrm{SNR}_{p}[n]<\mathrm{SNR}_{\mathrm{th}}\), set  
        \begin{align}
          \hat{\mathbf{s}}_{\mathcal{T}}[n]=\hat{\mathbf{s}}_{\mathcal{T}}^{-}[n], \quad
          \mathbf C[n]=\mathbf C^{-}[n]
        \end{align}
        and skip the update.
      \item \emph{Update step (only when $\mathrm{SNR}_{p}[n]\;\ge\;\mathrm{SNR}_{\mathrm{th}}$)}:
      \begin{align}
        \mathbf{K}[n]
        &= \mathbf{C}^{-}[n]\mathbf{H}^{\!\mathsf{T}}
          \Bigl(\mathbf{H}\,\mathbf{C}^{-}[n]\mathbf{H}^{\!\mathsf{T}} + \mathbf{R}[n]\Bigr)^{-1}, \\
        \hat{\mathbf{s}}_{\mathcal{T}}[n]
        &= \hat{\mathbf{s}}_{\mathcal{T}}^{-}[n]
           + \mathbf{K}[n]\bigl(\mathbf{z}[n] - \mathbf{H}\,\hat{\mathbf{s}}_{\mathcal{T}}^{-}[n]\bigr), \\
        \mathbf{C}[n]
        &= \bigl(\mathbf{I}_4 - \mathbf{K}[n]\mathbf{H}\bigr)\,\mathbf{C}^{-}[n],
      \end{align}
    \end{itemize}
    where superscript ``\(-\)'' indicates a predicted value before new measurements, and \(\mathbf{K}[n]\) is the Kalman gain. Notice that although the radar directly measures only position, the filter infers velocity through~\eqref{eq:cv_model_short}.

    \subsection{Age of Information (AoI) Model}
    \label{sec:AoI}
    The AoI measures the timeliness of the target state updates received by the ground users. Let $\Delta_k[n]$ denote the AoI for user $k$ at the end of slot $n$. In simple terms, $\Delta_k[n]$ counts the number of slots since the latest update (i.e., the target state sensed by the UAV) was generated and successfully received by user $k$. Let $g[n]$ be the slot in which the most recent target state was generated up to and including slot $n$. For slots $n \ge 2$, $g[n]$ is updated according to
    
    \begin{equation}
    g[n] =
    \begin{cases}
      n, & \text{if }     \mathrm{SNR}_{p}[n]\;\ge\;\mathrm{SNR}_{\mathrm{th}},\\[2mm]
      g[n-1], & \text{otherwise},
    \end{cases}
    \label{eq:update_generation}
    \end{equation}
    with the initial condition $g[1] = 1$. The AoI of user $k$ evolves as follows. For $n \ge 2$, we have:
    \begin{equation}
    \Delta_k[n] =
    \begin{cases}
      n - g[n-1], & \text{if } \Gamma_k[n] \ge \Gamma_{\text{th}},\\[2mm]
      \Delta_k[n-1] + 1, & \text{otherwise},
    \end{cases}
    \label{eq:aoi_update}
    \end{equation}
    with the initial condition $\Delta_k[1] = 1$. If user $k$ successfully decodes the update (i.e., its signal quality $\Gamma_k[n] \ge \Gamma_{\text{th}}$), the AoI resets to the elapsed time since the update was generated. If the decoding fails, the AoI simply increases by one.

    The average AoI of all users at time slot $n$ is denoted as 
    \begin{equation}
        \bar{\Delta}[n] = \frac{1}{K}\sum_{k=1}^{K} \Delta_k[n].
    \end{equation}
    Over the entire horizon of $N$ slots, we define the the long-term time-averaged AoI across all $K$ users is given by
    \begin{equation}
    \bar{\Delta} = \frac{1}{N}\sum_{n=1}^{N} \bar{\Delta}[n]
    = \frac{1}{NK}\sum_{n=1}^{N}\sum_{k=1}^{K} \Delta_k[n] .
    \end{equation}

    \subsection{Problem Formulation}

    Our objective is to jointly design the UAV’s trajectory and transmit beamforming to minimize the long-term average AoI at the ground users. The design variables are
    \begin{equation}
        \{\mathbf{p}_u[n],\, \{\mathbf{w}_k[n]\}_{k\in\mathcal{K}},\, \mathbf{w}_{\mathcal{T}}[n]\},
        \label{eq:optimize_variables}
    \end{equation}
    where \(\mathbf{p}_u[n]\) is the UAV’s position at time slot \(n\), \(\mathbf{w}_k[n]\) is the communication beamforming vector for user \(k\) at slot \(n\), and \(\mathbf{w}_{\mathcal{T}}[n]\) is the sensing beamform vector.
    The optimization problem is formulated as
    \begin{equation}
    \begin{aligned}
    & \textbf{P1}: \ 
    \min_{\{\mathbf{p}_u[n],\{\mathbf{w}_k[n]\}_{k\in\mathcal{K}},\mathbf{w}_{\mathcal{T}}[n]\}} \bar{\Delta} = \frac{1}{KN} \sum_{k=1}^{K} \sum_{n=1}^{N} \Delta_k[n] \\[2mm]
    \textrm{s.t.}\quad 
    & \text{C1:} \ \sum_{k=1}^{K}\|\mathbf{w}_k[n]\|^2 + \|\mathbf{w}_{\mathcal{T}}[n]\|^2 \le P_{\max}, \\[2mm]
    & \text{C2:} \ \|\mathbf{p}_u[n+1]-\mathbf{p}_u[n]\| \le v_{\max}\,\delta_t, \\
    & \text{C3:} \  \eqref{eq:update_generation} \ \text{and} \ \eqref{eq:aoi_update},
    \end{aligned}
    \label{eq:problem}
    \end{equation}
     where constraint C1 enforces the per-slot transmit power limit, C2 regulates the UAV’s mobility and C3 defines the AoI updated rule.

\section{Proposed DRL-based Joint UAV Trajectory and Beam Control}
\label{sec:Proposed_Solution}

This section presents a novel DRL-based approach to address the joint UAV trajectory and beamforming optimization problem formulated in \eqref{eq:problem}. First, the overall solution architecture is outlined. Then, the joint optimization problem is reformulated as a DRL-driven sequential decision-making process with well-defined state, action, and reward structures. Subsequently, we introduce post-processing modules for power allocation and beam synthesis based on the DRL output. Finally, the complete SAC-based training pipeline is detailed.

\subsection{Solution Architecture}

\begin{figure*}
        \centering
        \includegraphics[width=1\textwidth]{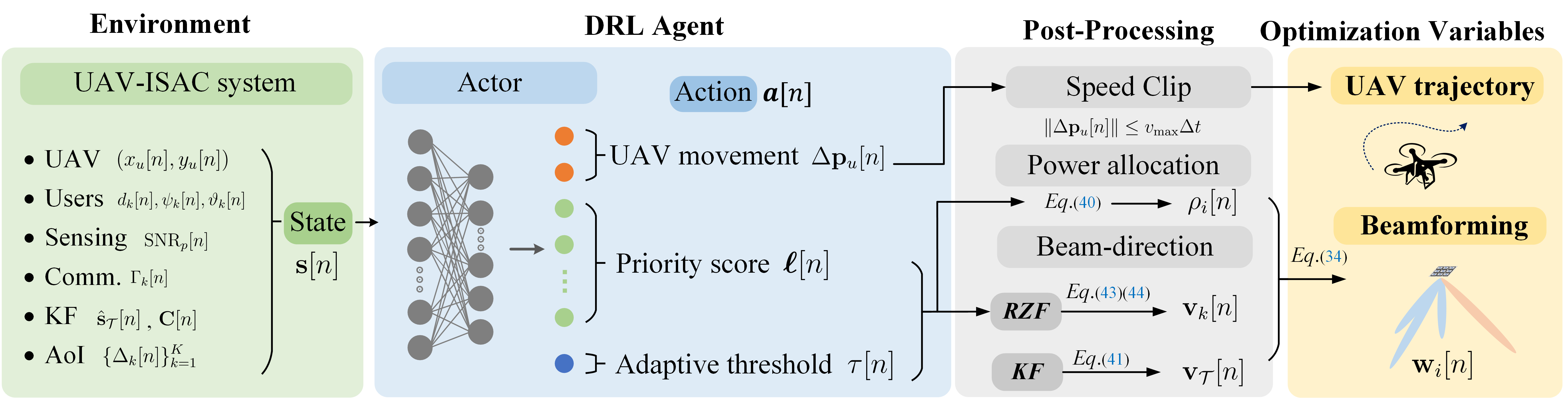}
        \caption{One-step decision making of the optimization variables based on DRL.}
        \label{fig:rl}
    \end{figure*}

    We employ DRL to jointly optimize the variables defined in~\eqref{eq:optimize_variables}. Fig.~\ref{fig:rl} illustrates the one-step decision making of the UAV trajectory and beamforming: (i) Environment observes state; (ii) DRL agent outputs UAV motion, priority scores, and adaptive threshold; and (iii) Post-processing maps to power/beam and forms the final transmit vector. 

    For the UAV trajectory, the DRL agent directly outputs the UAV’s next position at each time slot, while ensuring compliance with the velocity constraint.
    
    For the beamforming, we reformulate the beamforming vector in~\eqref{eq:beam} as:
    \begin{equation}
        \mathbf{w}_i[n] \;=\; \sqrt{\rho_i[n] \, P_{\max}}\;{\mathbf{v}}_i[n],
        \quad
        i \in \{\mathcal{T},1,\ldots,K\},
        \label{eq:beam_decompose}
    \end{equation}
    where \(\rho_i[n] \in [0,1]\) denotes the power allocation ratio for user or target \(i\), subject to \(\sum_{i} \rho_i[n] = 1\), and \(\mathbf{v}_i[n]\) represents the beam direction vector. As such, the beamforming vector is constructed by jointly determining both the power allocation and direction for each beam.
    
    For beam power allocation, due to the practical constraints in UAV-ISAC systems such as limited transmission power, finite antenna array size, and UAV position, not all users or targets can be effectively served in every time slot. Therefore, the DRL agent outputs a set of priority scores \(\boldsymbol{\ell}[n]\) for all beams (covering both the target and the users), along with an adaptive threshold \(\tau\), to guide the allocation of power shares. The detailed power allocation strategy is described in Section~\ref{subsec:power}.
    
    For beam direction control, we exploit the known locations of communication users and apply regularized zero forcing (RZF) to mitigate inter-user interference. For the sensing target, whose position is unknown and dynamic, we leverage KF to predict its location and direct the sensing beam accordingly. The beam direction design methodology is detailed in Section~\ref{subsec:beam}.

\subsection{DRL-Based Decision Making}
    The joint trajectory–beamforming problem \textbf{(P1)} seeks to minimize the long-term average AoI subject to UAV mobility constraints, per-slot power limits, strong space–time coupling among users and beams, and stochastic target dynamics. Due to its high dimensionality, non-convexity, and partially unknown environment, direct optimization is impractical. We therefore recast it as a finite-horizon MDP \cite{puterman2014markov}, enabling a learning-driven solution via DRL.

    \subsubsection{From Optimization \textbf{(P1)} to RL Reformulation}
    RL treats the controller as an agent that continuously interacts with an environment. The agent–environment interaction can be mathematically formulated as a finite MDP, defined by the tuple
    \(\mathcal{M}=\langle\mathcal{S},\mathcal{A},\mathcal{P},
               \mathcal{R},\gamma,N\rangle\),
    where \(\mathcal{S}\) and \(\mathcal{A}\) are the state and action spaces, \(\mathcal{P}: \mathcal{S}\!\times\!\mathcal{A} \rightarrow\mathcal{P}(\mathcal{S})\) is the transition kernel \(p(\mathbf{s}[n+1]\mid\mathbf{s}[n],\mathbf{a}[n])\),
    \(\mathcal{R}: \mathcal{S}\!\times\!\mathcal{A}\rightarrow\mathbb{R}\)
    is the instantaneous reward,
    \(\gamma\in(0,1]\) the discount factor,
    and \(N\) the episode length (same as the number of time slots in the system model).
    At decision epoch \(n\), the DRL agent observes \(\mathbf{s}[n]\in\mathcal{S}\),
    executes action \(\mathbf{a}[n]\in\mathcal{A}\),
    receives reward \(r[n]=\mathcal{R}(\mathbf{s}[n],\mathbf{a}[n])\),
    and the environment draws
    \(\mathbf{s}[n+1]\sim\mathcal{P}(\cdot\mid\mathbf{s}[n],\mathbf{a}[n])\).
    The agent seeks a policy
    \(\pi:\mathcal{S}\rightarrow\mathcal{P}(\mathcal{A})\)
    that maximises the expected return
    \(\sum_{n=1}^{N}\gamma^{\,n-1}r[n]\).
    
    In our UAV-ISAC system, the environment comprises the UAV-enabled ISAC system, including the elements such as the UAV, the $K$ ground users, and the sensing target, and also the communication and sensing processing, and the AoI updating rule. The agent is the decision maker (a neural network policy~$\pi_{\boldsymbol\theta}$ discussed in Section~\ref{subsec:sac_training}) to output trajectory and beamforming strategy.
    The following subsections detail the designed MDP.

    \subsubsection{State~$\mathbf{s}[n]$}
    
    The state vector $\mathbf{s}[n]$ serves as a minimal sufficient statistic for one-step decision-making by the UAV. It compactly encodes all relevant mobility, communication, and sensing information at slot $n$:
    \begin{align}
    \mathbf s[n]=\bigl[
    &\underbrace{x_u[n],y_u[n]}_{\text{UAV}},\;
     \underbrace{\{d_k[n],\psi_k[n],\vartheta_k[n],\Gamma_k[n],\Delta_k[n]\}_{k=1}^{K}}_{\text{Users}},\nonumber\\
    &\underbrace{\hat{\mathbf{s}}_{\mathcal{T}}[n], \mathrm{SNR}_{p}[n]}_{\text{Target}},\;
     \underbrace{\operatorname{diag}(\mathbf C[n]),\operatorname{tr}(\mathbf C[n])}_{\text{KF Uncertainty}},\;
     \underbrace{\bar\Delta[n],\varrho[n]}_{\text{Global}}
    \bigr],
    \label{eq:state_vec}
    \end{align}
    where the individual components are organized into five functional groups:
    \begin{itemize}
      \item UAV: the UAV’s horizontal position $(x_u[n],y_u[n])$;
      \item Users: users' relative geometry $(d_k[n],\psi_k[n],\vartheta_k[n])$, instantaneous downlink $\mathrm{SINR}$ $\Gamma_k[n]$, and AoI $\Delta_k[n]$, jointly reflecting link quality and service urgency;
      \item Target: the KF estimate $\hat{\mathbf{s}}_{\mathcal T}[n]$ and the pulse-compressed radar $\mathrm{SNR}_{p}[n]$, indicating sensing accuracy and echo strength;
      \item KF Uncertainty: The element-wise variance $\operatorname{diag}\bigl(\mathbf C[n]\bigr)$ and total variance $\operatorname{tr}\bigl(\mathbf C[n]\bigr)$ derived from the covariance matrix $\mathbf{C}[n]$, which together quantify estimation uncertainty;
      \item Global: the average AoI $\bar\Delta[n]$ and the normalized progress index $\varrho[n]\triangleq n/N$ provide mission-level temporal context.
    \end{itemize}
    This structured representation enables the DRL agent to access all relevant local and global context required for informed trajectory and beamforming decisions.

    \subsubsection{Action~$\mathbf{a}[n]$}
    Rather than learning a full complex‐valued beam vector, we output 
    \begin{equation}
        \mathbf{a}[n] = \bigl[\,\Delta\mathbf{p}_u[n],\,\boldsymbol{\ell}[n],\,\tau[n]\bigr].
        \label{eq:action_output}
    \end{equation}

    \begin{itemize}
    \item $\Delta\mathbf{p}_u[n]$ directly controls the two‐dimensional UAV displacement and is later clipped to respect the velocity limit $\|\Delta\mathbf{p}_u[n]\|\le v_{\max}\Delta t$;
    \item the priority scores $\boldsymbol{\ell}[n]$ and adaptive threshold $\tau[n]$ compactly encode power‐allocation intentions and are mapped to a feasible power simplex $\{\rho_i[n]\}$ via~\eqref{eq:rho_softmax}.
    \end{itemize}

    \subsubsection{Transition kernel \(\mathcal{P}\)}  
    Given \((\mathbf{s}[n],\mathbf{a}[n])\), the next state
    \(\mathbf{s}[n{+}1]\) is obtained by propagating the current variables through the physical models specified in
    Section~\ref{sec:System_Model}, such as UAV and target movement, users' SINR calculation, sensing measurement, KF estimation, and AoI updating.

    \subsubsection{Reward Function}
    The RL agent is trained to minimize the network-wide AoI, which is equivalently realized by maximizing the negative AoI:
    \begin{equation}
        r[n] \;=\; -\bar{\Delta}[n].
        \label{eq:reward_function}
    \end{equation}
    Over an episode of $N$ slots, the agent therefore maximises the expected
    return $\sum_{n=1}^{N}\gamma^{\,n-1}r[n]$ with discount factor
    \mbox{$\gamma\in(0,1)$}, leading to UAV trajectories and beam-control actions that jointly keep user information fresh and the sensing beam on target,
    thus solving the optimization problem in~(\ref{eq:problem}).

\subsection{Power Allocation and Beam Design}
\label{subsec:power_beam_design}

At each slot~\(n\), the DRL agent outputs the continuous action vector
\(\mathbf a[n]=[\Delta\mathbf p_u[n],\,\boldsymbol{\ell}[n],\,\tau[n]]\).
The components \(\boldsymbol{\ell}[n]\) and threshold \(\tau[n]\) jointly determine the power allocation and beam directions for the sensing task and user communication, as described below.

\subsubsection{Power allocation}
\label{subsec:power}
The sensing beam (target beam) is always scheduled, independent of the threshold. The set of scheduled users is explicitly determined by comparing each user's logit \(\ell_k[n]\) with the threshold \(\tau[n]\)
\begin{equation}
    \mathcal U[n]=\{k\mid\ell_k[n]\ge\tau[n]\}.
\end{equation}
To avoid the degenerate scenario where no user is scheduled, if \(\mathcal U[n]\) is empty, the user with the highest logit is selected by
\begin{equation}
    \mathcal U[n]=\{\arg\max_k\ell_k[n]\},\quad\text{if}\quad|\mathcal U[n]|=0.
\end{equation}
Applying a softmax function to the logits yields the power-splitting ratios for all beams as
\begin{equation}
\label{eq:rho_softmax}
    \rho_i[n] =
    \begin{cases}
        \displaystyle\frac{\exp(\ell_i[n])}{\sum_{j\in \{\mathcal T\}\cup \mathcal U[n]} \exp(\ell_j[n])},
          & i \in \{\mathcal T\}\cup \mathcal U[n], \\[12pt]
        0, & \text{otherwise}.
    \end{cases}
\end{equation}

\subsubsection{Beam-direction synthesis}
\label{subsec:beam}
The sensing beam direction is computed directly using angles predicted by the KF, denoted as a function of \((\psi_{\mathcal{T}}[n],\vartheta_{\mathcal T}[n])\)
\begin{equation}
\label{eq:v_target}
\mathbf v_{\mathcal T}[n]=
\frac{\mathbf a(\psi_{\mathcal{T}}[n],\vartheta_{\mathcal T}[n])}
{\|\mathbf a(\psi_{\mathcal{T}}[n],\vartheta_{\mathcal T}[n])\|_2}.
\end{equation}

To generate beamforming directions for scheduled users, we adopt the RZF approach. Let
\(\mathbf H_{\mathcal S}[n]\in\mathbb C^{|\mathcal U[n]|\times M}\) 
collect their conjugate-transpose channel vectors. Define the adaptive regularization factor as
\begin{equation}
    \alpha[n]=\max\left(10^{-9},\;\frac{|\mathcal U[n]|\xi_k^2}{\sum_{k\in\mathcal U[n]}P_k[n]}\right),
\end{equation}
where $\xi_k^2$ is the receiver noise variance. The unnormalized beamforming matrix is
\begin{equation}
\label{eq:rzf_matrix}
    \widetilde{\mathbf V}_{\mathcal S}[n]=
    \mathbf H_{\mathcal S}^{\mathrm H}[n]
    \bigl(\mathbf H_{\mathcal S}[n]\mathbf H_{\mathcal S}^{\mathrm H}[n]+\alpha[n]\mathbf I\bigr)^{-1}.
\end{equation}
where $\mathbf I$ denotes the identity matrix. Each user's beam direction is then normalized as
\begin{equation}
\label{eq:v_user}
\mathbf v_k[n]=
\frac{\widetilde{\mathbf v}_k[n]}
{\|\widetilde{\mathbf v}_k[n]\|_2},\quad k\in\mathcal U[n],
\end{equation}
where \(\widetilde{\mathbf v}_k[n]\) is the corresponding column of \(\widetilde{\mathbf V}_{\mathcal S}[n]\). The beamforming vectors for the remaining unscheduled users (\(k\notin\mathcal U[n]\)) are explicitly set to zero vectors. Thus, the overall beamforming matrix consists of the sensing beam vector, scheduled users' beam vectors, and zero vectors for unscheduled users.

\subsection{SAC-based Agent Training}
\label{subsec:sac_training}

\begin{algorithm}[!t]
\footnotesize
\SetCommentSty{small}
\LinesNumbered
\caption{SAC-Based Joint UAV Trajectory and Beam Control}
\label{alg:SAC_UAV_ISAC}
\KwIn{$N$, $\gamma$, Training max episodes $E$, initial temperature $\kappa_0$, $\mathcal{H}_{\text{tar}}$, learning rates $\eta_\theta,\eta_\phi,\eta_\kappa$, mini-batch size $\mathcal{D}$, update-interval $I_{\mathrm{update}}$, grad-repeat $\mathcal{J}$}
\KwOut{Trained actor parameters $\boldsymbol\theta^{*}$} 

\textbf{Initialization:}\\
\Indp
Initialise actor $\pi_{\boldsymbol\theta}$, critics $Q_{\boldsymbol\phi_1},Q_{\boldsymbol\phi_2}$,\\
\hspace*{1em}target critics $\hat Q_{\bar{\boldsymbol\phi}_1},\hat Q_{\bar{\boldsymbol\phi}_2}\!\leftarrow Q_{\boldsymbol\phi_1},Q_{\boldsymbol\phi_2}$;\\
Initialise temperature $\kappa\!\leftarrow\!\kappa_0$; replay buffer $\mathcal B\!\leftarrow\!\varnothing$.\\
\Indm

\For{$e\!=\!1$ \KwTo $E$}{
    Reset environment; obtain initial state $\mathbf s[1]$; $R_e\!\leftarrow\!0$.\\

    \For{$n\!=\!1$ \KwTo $N$}{       
        Sample action $\mathbf a[n]\!\sim\!\pi_{\boldsymbol\theta}(\,\cdot\!\mid\!\mathbf s[n])$\\
        
        \  \textbf{UAV position:} $\mathbf p_u[n\!+\!1]=\mathbf p_u[n]+\Delta\mathbf p_u[n]$.\\
        
        \quad \quad  Power allocation: $\{\rho_i[n]\}$ via~(\ref{eq:rho_softmax}).\\
        \quad \quad Sensing beam: $\mathbf v_{\mathcal T}[n]$ by KF estimation~(\ref{eq:v_target});\\
        \quad \quad User beams: $\mathbf v_k[n]$ via RZF~(\ref{eq:rzf_matrix})–(\ref{eq:v_user}).\\
        
        \  \textbf{Beamforming vector:} $\mathbf w_i[n]$ \eqref{eq:beam_decompose};\\

        Environment Interaction: \\ 
        \quad \quad Reward $r[n]=-{\textstyle\sum_{k}}\Delta_k[n]$;\\
        \quad \quad Next state $\mathbf s[n+1]$, set $\text{done}\leftarrow(n==N)$.\\
        
        Store $(\mathbf s[n],\mathbf a[n],r[n],\mathbf s[n\!+\!1],\text{done})$ in $\mathcal B$;\\
        $R_e\!\leftarrow\!R_e + r[n]$.\\
        
        \If{$|\mathcal{B}|\ge \mathcal{D}$ \textbf{and} $n \bmod I_{\mathrm{update}} = 0$}{ 
           \For{$j=1$ \KwTo $\mathcal{J}$}{
              Sample mini-batch $\mathcal{D}$ from $\mathcal{B}$; \\
              Compute $y[n]$ \eqref{eq:td_target_time}, $\mathcal{L}_Q$ \eqref{eq:critic_loss_time}, $\mathcal{L}_\pi$ \eqref{eq:actor_loss_time}, $\mathcal{L}_\kappa$ \eqref{eq:temp_loss_general};  
           }
        Gradient steps for actor, critic, temperature~ \eqref{eq:gradient_updates} 

        Soft update for target net~\eqref{eq:soft}.}
    }
}
\Return{$\boldsymbol\theta^{*}$}
\end{algorithm}
This subsection describes how the agent is trained to obtain an effective policy using the SAC \cite{haarnoja2018soft}. We first outline the actor–critic (AC) structure, then highlight SAC’s advantages, and finally detail the network design and update rules.

\subsubsection{Overview and Rationale}
AC methods employ a policy network (actor) for action selection and a value network (critic) to evaluate these actions. SAC is a variant of AC, which is is well-suited for UAV-ISAC tasks due to its training stability and efficient exploration. It employs two critics to reduce overestimation bias and incorporates entropy regularization with automatic temperature adjustment, enabling balanced exploration–exploitation. This facilitates more accurate trajectory and beamforming policies over time.

\subsubsection{Neural-Network Architectures}
Both the actor and critics are parameterized by neural networks. Below are the key components:

\paragraph{Actor \(\pi_{\boldsymbol\theta}\)}
We maintain a parameterized policy \(\pi_{\boldsymbol\theta}\) that outputs a distribution over \(\mathbf{a}[n]\) given \(\mathbf{s}[n]\). In our case, \(\pi_{\boldsymbol\theta}\) is a neural network producing the mean \(\boldsymbol{\mu}(\mathbf{s}[n])\) and log-standard-deviation \(\log\boldsymbol{\sigma}(\mathbf{s}[n])\) of a multivariate Gaussian, from which we sample an action and map it (e.g., via \(\tanh\)) into a valid control range. Through training, the parameters $\boldsymbol\theta$ are optimized to maximize the expected return, and the final policy is denoted by the optimal parameters $\boldsymbol\theta^*$.

\paragraph{Twin Critics \(Q_{\boldsymbol\phi_1}, Q_{\boldsymbol\phi_2}\) \& Target Critics}
Each critic \(Q_{\boldsymbol\phi_i}\) approximates the action-value function
\begin{equation}
  Q^{\pi}(\mathbf{s}[n], \mathbf{a}[n])
  \;=\;
  \mathbb{E}_{\pi}\!\biggl[
    \sum_{k=0}^{\infty}
    \gamma^{\,k}\,r[n+k]
    \;\Big|\;\mathbf{s}[n], \mathbf{a}[n]
  \biggr],
  \label{eq:action_value_definition}
\end{equation}
In SAC, we use two critics, \(Q_{\boldsymbol\phi_1}\) and \(Q_{\boldsymbol\phi_2}\), to reduce overestimation. Each critic is a neural network mapping \((\mathbf{s}[n], \mathbf{a}[n])\) to a scalar \(Q\)-value. To stabilize learning, we maintain an additional set of slowly-updated copies
\(\{\hat{Q}_{\bar{\boldsymbol\phi}_1},\hat{Q}_{\bar{\boldsymbol\phi}_2}\}\),
often called target critics.  
They are initialized with \(\bar{\boldsymbol\phi}_i\!=\!\boldsymbol\phi_i\) and kept
close to the online critics via a soft update rule described later in~\eqref{eq:soft}.

\subsubsection{SAC Training Mechanism} Algorithm~\ref{alg:SAC_UAV_ISAC} summarizes the complete SAC-based training procedure. SAC maximizes the following objective that combines reward and policy entropy:
\begin{equation}
  J_{\pi}
  \;=\;
  \mathbb{E}\!\biggl[
    \sum_{n=1}^{N}
    \gamma^{\,n-1}
    \Bigl(
      r[n]
      \;+\;
      \kappa\,\mathcal{H}(\pi(\mathbf{a}[n]\mid\mathbf{s}[n]))
    \Bigr)
  \biggr],
  \label{eq:sac_objective}
\end{equation}
where \(\kappa>0\) balances exploitation (maximizing reward) and exploration (maximizing entropy \(\mathcal{H}(\cdot)\)).

\paragraph{Critics Update}
When training the critics, we form a Temporal-Difference (TD) target using a mini-batch of transitions
\(\bigl(\mathbf{s}[n], \mathbf{a}[n], r[n], \mathbf{s}[n+1], \text{done}\bigr)\)
sampled from a replay buffer, where $\text{done}$ indicates whether state $\mathbf{s}[n+1]$ is terminal (end of episode)
\begin{equation}
    \begin{aligned}
    y[n] = r[n] + \gamma\,&(1-\mathrm{done}) \Bigl[
    \min_{j}\,
    \hat{Q}_{\bar{\boldsymbol\phi}_j}\bigl(\mathbf{s}[n+1], \mathbf{a}[n+1]\bigr)\\
    & -
    \kappa\,\log\pi_{\boldsymbol\theta}\bigl(\mathbf{a}[n+1]\mid\mathbf{s}[n+1]\bigr)
  \Bigr],
\end{aligned}    
  \label{eq:td_target_time}
\end{equation}
where \(\mathbf{a}[n+1]\) is drawn from the current policy
\(\pi_{\boldsymbol\theta}(\mathbf{s}[n+1])\), and
\(\hat{Q}_{\bar{\boldsymbol\phi}_j}\) are target networks (softly updated copies of \(Q_{\boldsymbol\phi_j}\)).
Each critic \(Q_{\boldsymbol\phi_i}\) is updated by minimizing
\begin{equation}
  \mathcal{L}_{Q}
  \;=\;
  \tfrac12
  \sum_{i=1}^{2}
  \mathbb{E}_{\mathcal{B}}\Bigl[
    \bigl(
      Q_{\boldsymbol\phi_i}(\mathbf{s}[n], \mathbf{a}[n]) \;-\; y[n]
    \bigr)^2
  \Bigr].
  \label{eq:critic_loss_time}
\end{equation}

\paragraph{Actor Update}
The actor is then updated to maximize the Q-value and maintain high entropy. Specifically, we minimize
\begin{equation}
\begin{aligned}
      \mathcal{L}_{\pi}
  \;=\;
  \mathbb{E}_{\mathcal{B}}\Bigl[
    \kappa\,&\log\pi_{\boldsymbol\theta}\bigl(\mathbf{a}[n]\mid\mathbf{s}[n]\bigr)
    \;-\;
    Q_{\boldsymbol\phi_1}\bigl(\mathbf{s}[n], \mathbf{a}[n]\bigr)
  \Bigr],
  \quad \\
  &\mathbf{a}[n]\sim\pi_{\boldsymbol\theta}(\mathbf{s}[n]).
  \end{aligned}
  \label{eq:actor_loss_time}
\end{equation}

\paragraph{Temperature Tuning}
We also learn \(\kappa\) online to meet a target entropy \(\mathcal{H}_{\mathrm{tar}}\)
\begin{equation}
  \mathcal{L}_{\kappa}
  \;=\;
  \mathbb{E}_{\mathcal{B}}\Bigl[
    \kappa\,
    \bigl(
      -\log\pi_{\boldsymbol\theta}\bigl(\mathbf{a}[n]\mid\mathbf{s}[n]\bigr)
      \;-\; 
      \mathcal{H}_{\mathrm{tar}}
    \bigr)
  \Bigr].
  \label{eq:temp_loss_general}
\end{equation}

\subsubsection{Replay Buffer and Update Frequency}
To stabilize training and improve sample efficiency, we employ an experience replay buffer $\mathcal{B}$ that stores transitions \mbox{$(\mathbf{s}[n], \mathbf{a}[n], r[n], \mathbf{s}[n+1], \text{done})$}. The replay buffer reduces temporal correlations by uniformly sampling from a large memory pool. Parameter updates occur every $I_{\mathrm{update}}$ steps once the buffer contains at least $\mathcal{D}$ transitions. Each update involves sampling a mini-batch of size $\mathcal{D}$ and repeating gradient descent steps $\mathcal{J}$ times to enhance learning stability.

\subsubsection{Gradient-Based Parameter Updates}
Given the previously defined loss functions (see~\eqref{eq:critic_loss_time},~\eqref{eq:actor_loss_time}, and~\eqref{eq:temp_loss_general}), the gradient-based parameter updates at each training iteration are explicitly formulated as follows:
\begin{equation}
\label{eq:gradient_updates}
\begin{aligned}
    \boldsymbol{\phi}_i &\leftarrow \boldsymbol{\phi}_i - \eta_{\phi}\,\nabla_{\boldsymbol{\phi}_i}\,\mathcal{L}_{Q}, \quad i=1,2,\\
    \boldsymbol{\theta} &\leftarrow \boldsymbol{\theta} - \eta_{\theta}\,\nabla_{\boldsymbol{\theta}}\,\mathcal{L}_{\pi},\\
    \kappa &\leftarrow \kappa - \eta_{\kappa}\,\nabla_{\kappa}\,\mathcal{L}_{\kappa}.
\end{aligned}
\end{equation}
where $\eta_\theta$, $\eta_\phi$, and $\eta_\kappa$ are the learning rates for the actor, critic, and entropy temperature $\kappa$, respectively.

The target critics \(\hat{Q}_{\bar{\boldsymbol\phi}_i}\) are slowly synchronized with the primary critics:
\begin{equation}
\label{eq:soft}
  \bar{\boldsymbol\phi}_i
  \;\leftarrow\;
  \eta_s\boldsymbol\phi_i
  \;+\;
  (1-\eta_s)\,\bar{\boldsymbol\phi}_i,
\end{equation}
where $\eta_s$ is the soft-update coefficient, typically small to ensure stable TD targets.

\subsubsection{Complexity Analysis}
The Markov state and action dimensions are $d_{\mathrm{s}} = 5K + 14$ and $d_{\mathrm{a}} = K + 3$. A two-layer fully connected actor network with hidden width $N_{\mathrm{h}}$ incurs computational cost $F_{\pi} = d_{\mathrm{s}}N_{\mathrm{h}} + N_{\mathrm{h}}^2 + 2d_{\mathrm{a}}N_{\mathrm{h}}$ per forward pass, while a single critic update requires $F_{Q} = (d_{\mathrm{s}} + d_{\mathrm{a}})N_{\mathrm{h}} + N_{\mathrm{h}}^2 + N_{\mathrm{h}}$. Since SAC maintains two critics, each update step consumes $2F_{Q}$. During training, an episode of $N$ time slots involves (i) environment interactions with per-slot cost $\mathcal{O}(F_{\pi} + K + M^3)$ and (ii) network updates every $I_{\mathrm{upd}}$ slots, each repeated $J$ times on mini-batches of size $D$. The total training complexity per episode is $\mathcal{O}\big(NM^3 + \frac{NJD}{I_{\mathrm{upd}}}(F_{\pi} + 2F_{Q})\big)$.

After training, the agent performs inference at each decision slot. This includes actor evaluation ($\mathcal{O}(F_{\pi})$), Softmax-based power allocation ($\mathcal{O}(K)$), and regularized zero-forcing beamforming ($\mathcal{O}(M^3)$), resulting in a per-slot inference complexity of $\mathcal{O}(F_{\pi} + K + M^3)$.

\section{Performance Evaluation}
\label{sec:SimulationResults} 
    
    \begin{table}[t]
    \centering
    \caption{Default parameters in simulation.}
    \label{table:parameters}
    \begin{tabular}{lll}
    \toprule
    Symbol & Physical meaning & Default Value \\ 
    \midrule
    $K$ & Number of ground users & $6$ \\
    $N$ & Number of time-slots & $60$ \\
    $\delta_t$ & Slot duration & $1\,\mathrm{s}$ \\
    $v_{\max}$ & UAV max horizontal speed & $20\,\mathrm{m/s}$ \\
    $H$ & UAV altitude & $50\,\mathrm{m}$ \\
    $M_x$ & UPA elements (x–axis) & $4$ \\
    $M_y$ & UPA elements (y–axis) & $4$ \\
    $f_c$ & Carrier frequency & $2\,\mathrm{GHz}$ \\
    $P_{\max}$ & Max transmit power & 20 $\mathrm{dBm}$ \\ 
    $G_{\mathrm{elem}}$ & Per-element antenna gain & 3~dBi \\
    $G_{\mathrm{user}}$ & User antenna gain & 0~dBi \\
    $\xi_k^2$ & Receiver noise power & $-90\ \mathrm{dBm}$ \\
    $\Gamma_{\mathrm{th}}$ & SINR decoding threshold & $10 \ \mathrm{dB}$ \\
    $\sigma_{\mathcal T}$ & Target radar cross-section & $1\,\mathrm{m^{2}}$ \\
    $T_{0}$ & System temperature & $290\,\mathrm{K}$ \\
    $B$ & Matched-filter bandwidth & $100\,\mathrm{MHz}$ \\
    $F$ & Receiver noise figure & $20\,\mathrm{dB}$ \\
    $N_{p}$ & Pulses per slot & $32$ \\
    $\sigma_{0}$ & Single-pulse range bound & $0.338\,\mathrm{m}$ \\
    $\sigma_{\mathrm{req}}$ & Required 1-$\sigma$ accuracy & $1\,\mathrm{m}$ \\
    $q_0^{2}$ & Process-noise variance & $0.25$ \\
    $V_{\max}^{\mathcal T}$ & Target max speed & $15\,\mathrm{m/s}$ \\
    $\gamma$ & Discount factor & 0.99 \\
    $\eta_\theta,\eta_\phi,\eta_\kappa$ & Learning rates & 0.0003\\
    $\eta_s$ & Soft-update coefficient & 0.01\\
    \bottomrule
    \end{tabular}
    \end{table}

\subsection{Simulation Setup}
The key simulation settings that govern the dynamics, initialization, and evaluation of the UAV-enabled ISAC system are summarized in Table~\ref{table:parameters}. Each simulation episode spans $N = 60$ discrete time slots, with a slot duration of $\delta_t = 1\,\mathrm{s}$. At the beginning of every episode, $K = 6$ ground users are uniformly and independently distributed within a 1,600 $\times$ 1,600,$\mathrm{m}^2$ square region. These users are considered to remain stationary throughout the episode. To ensure a fair comparison across different algorithms, we generate $100$ distinct random user layouts using seeds ranging from $100$ to $199$ and reuse these layouts across all schemes. The UAV starts each episode at a perturbed position near the initial location of the ground target, mimicking a realistic scenario in which the UAV is pre-deployed near the object of interest. Specifically, the initial UAV location is given by \mbox{$\mathbf{p}_u[0] = \mathbf{p}_0 + \Delta\mathbf{p}, \quad \Delta\mathbf{p} \sim \mathcal{N}(\mathbf{0},\,10^2\mathbf{I}_2)$}, where \mbox{$\mathbf{p}_0 = [350, 350]^{\mathsf T}\,\mathrm{m}$} is the target’s starting location, and $\Delta\mathbf{p}$ introduces small Gaussian noise in both horizontal dimensions.

    \subsubsection{Target Trajectory Model}

The target follows a randomly perturbed trajectory from the initial position $\mathbf{p}_0 = [350, 350]^{\mathsf{T}}\,\mathrm{m}$ to the destination \mbox{$\mathbf{p}_N = \text{[1,150, 1,150]}^{\mathsf{T}}\,\mathrm{m}$} over $N = 60$ slots. To ensure feasibility under the velocity constraint \mbox{$V_{\max}^{\mathcal{T}} = 15\,\mathrm{m/s}$}, we construct a hybrid path where the deterministic drift at slot $n$ is given by
\[
\mathbf{d}_{\mathrm{drift}}[n] = \frac{\mathbf{p}_N - \mathbf{p}_{\mathcal{T}}^{\mathrm{h}}[n]}{N - 1 - n},
\]
and the next position is set as
\[
\mathbf{p}_{\mathcal{T}}^{\mathrm{h}}[n{+}1] = \mathbf{p}_{\mathcal{T}}^{\mathrm{h}}[n] + \mathbf{d}_{\mathrm{drift}}[n] + \boldsymbol{\omega}[n],
\]
where $\boldsymbol{\omega}[n]$ is a random perturbation constrained to satisfy
\[
\|\boldsymbol{\omega}[n]\| \le \sqrt{(v_{\max}^{\mathcal{T}})^2 - \|\mathbf{d}_{\mathrm{drift}}[n]\|^2}.
\]
This construction guarantees the per-slot speed limit and ensures the terminal constraint $\mathbf{p}_{\mathcal{T}}^{\mathrm{h}}[N{-}1] = \mathbf{p}_N$ is met.

    \subsubsection{Compared Algorithms}
    
    To demonstrate the efficacy of the proposed SAC-based controller, we compare it against three baseline schemes:
    \begin{itemize}
        \item \textbf{Advantage Actor–Critic (A2C) \cite{mnih2016asynchronous}:} Shares the same state, action, and reward formulation as SAC but learns in an on-policy manner without entropy regularization.
        
        \item \textbf{Single-User AoI-Greedy Scheduling (SAGS):} Always selects the user with the largest instantaneous AoI and flies toward that user within the velocity constraint. The transmit power is equally split between sensing and the selected user beam.
        
        \item \textbf{Kalman-Forecast Random (KF-RAND):} Samples a random waypoint in a disc of radius $V_{\max}\delta_t$ centered at the Kalman-predicted target position, adds Gaussian jitter, and applies random user logits $\ell_k \sim \mathcal{N}(0, \sigma_{\text{logit}}^2)$. Beams with $\ell_k < 0$ are deactivated, and power is distributed via softmax among the remaining users.
    \end{itemize}
    
    All schemes are evaluated under identical conditions over 100 Monte Carlo episodes. This standardized setting enables reproducible and statistically sound performance comparisons.

\subsection{Simulation Results Analysis}

    \begin{figure}[t]
    \centering
    \includegraphics[width=0.9\columnwidth]{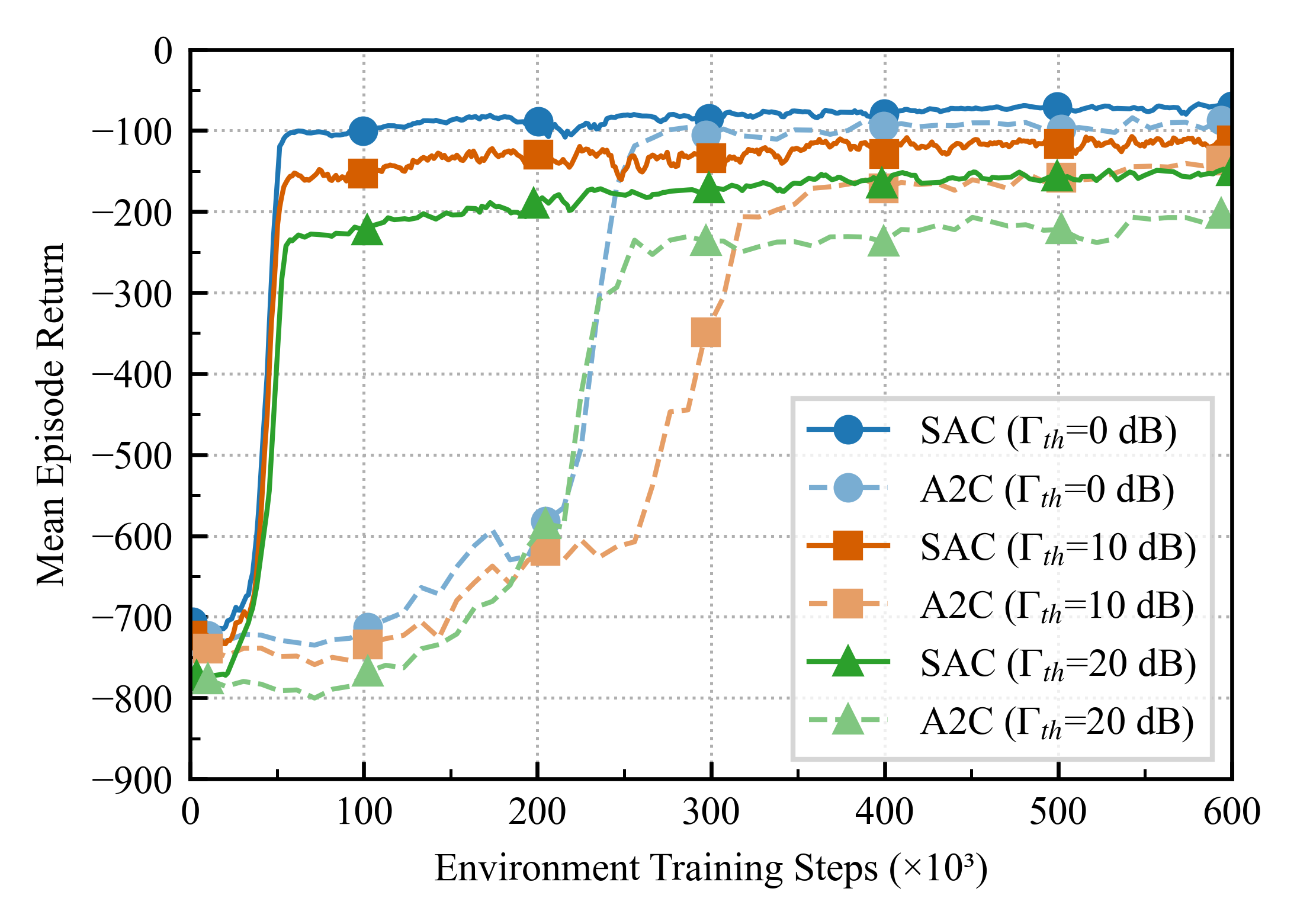}
    \caption{Training curves under different SINR thresholds $\Gamma_{\mathrm{th}}$.}
    \label{fig:training_curves_Gamma}
    \end{figure}

    \begin{figure}[t]
        \centering
        \includegraphics[width=0.9\columnwidth]{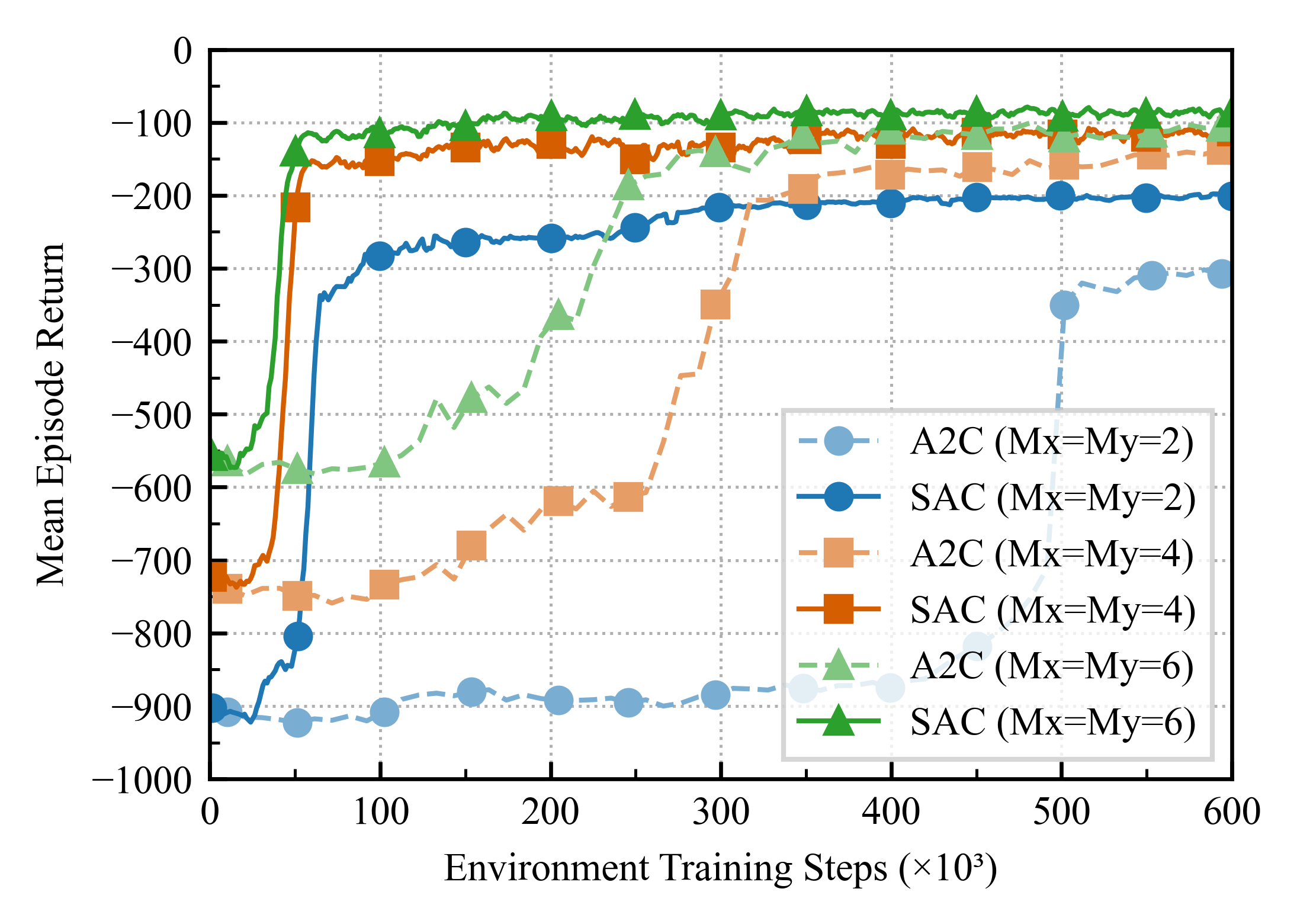}
        \caption{Training curves under different UPA configurations.}
        \label{fig:training_curves_UPA}
    \end{figure}
    Fig.~\ref{fig:training_curves_Gamma} and Fig.~\ref{fig:training_curves_UPA} compare the convergence performance of the proposed SAC algorithm and the A2C baseline under different system parameters. The horizontal axis denotes the total number of training steps, where each episode consists of N steps, i.e., 600,000 steps correspond to 10,000 episodes. The vertical axis represents the mean episode return, which is computed by averaging cumulative rewards per episode across 100 Monte Carlo runs under different user layouts. Overall, SAC consistently demonstrates faster convergence and greater training stability compared to A2C. This advantage arises from its off-policy update mechanism and entropy regularization, which together facilitate more efficient exploration and policy learning. In Fig.~\ref{fig:training_curves_Gamma}, a lower user SINR threshold $\Gamma_{\mathrm{th}}$ relaxes the communication quality requirements, which allows the agent to attain higher episode returns. As the threshold increases, the communication task becomes more challenging, resulting in moderate performance degradation for both algorithms. Nevertheless, SAC consistently outperforms A2C in all settings. In Fig.~\ref{fig:training_curves_UPA}, we evaluate the effect of antenna array size on convergence. Larger UPA sizes enhance the beamforming capability of the system, thereby improving both sensing and communication efficiency. This results in higher episode returns and faster convergence. The benefit is more pronounced for SAC, which is better able to exploit the increased system capacity. These results confirm that SAC not only offers superior learning stability but also scales more effectively with increasing system complexity.

    \begin{figure}[t]
    \centering
    \includegraphics[width=0.9\linewidth]{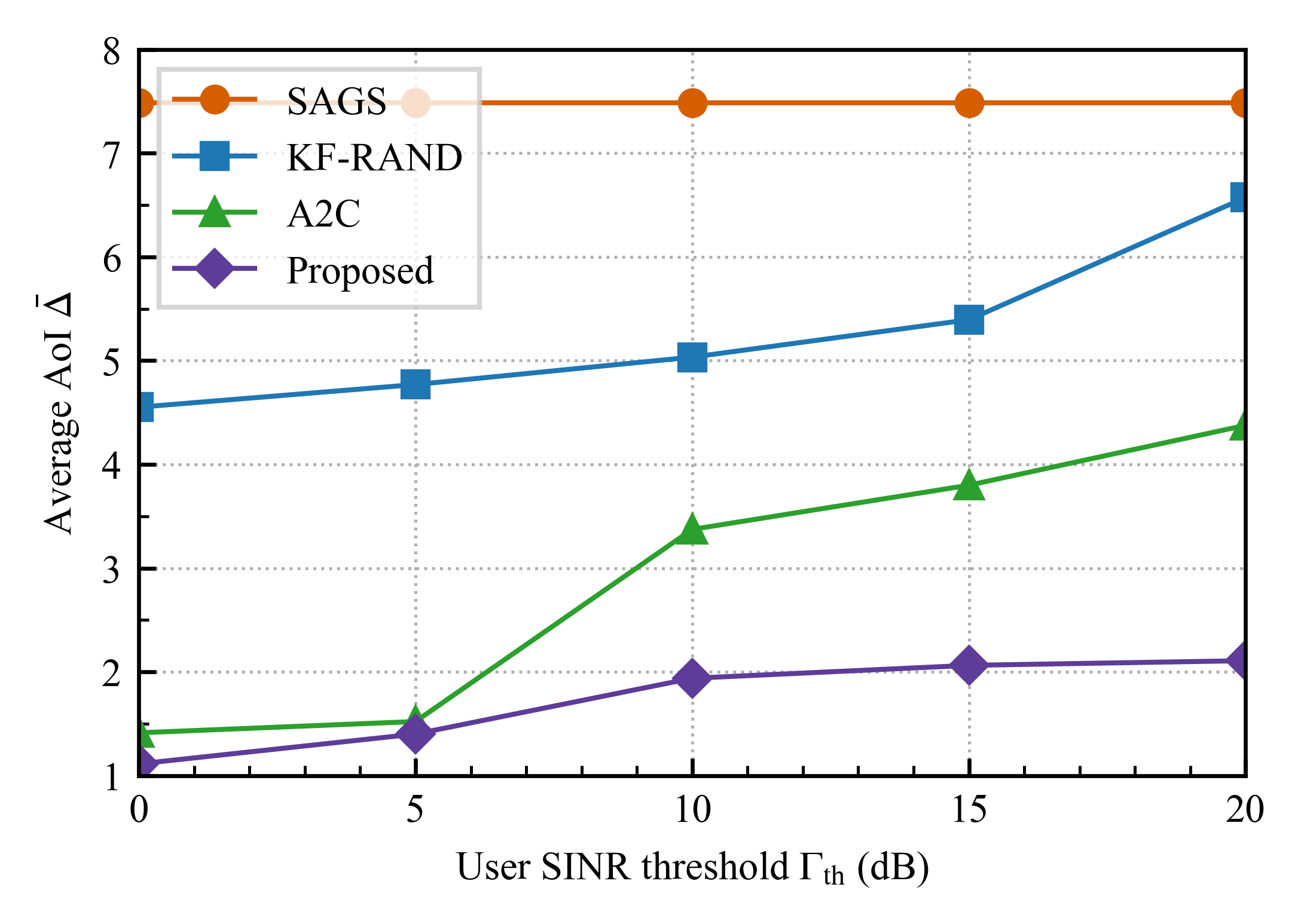}
    \caption{Average AoI versus user SINR threshold $\Gamma_{\mathrm{th}}$.}
    \label{fig:aoi_vs_sinr}
    \end{figure}
    
    \begin{figure}[t]
        \centering
        \includegraphics[width=0.9\columnwidth]{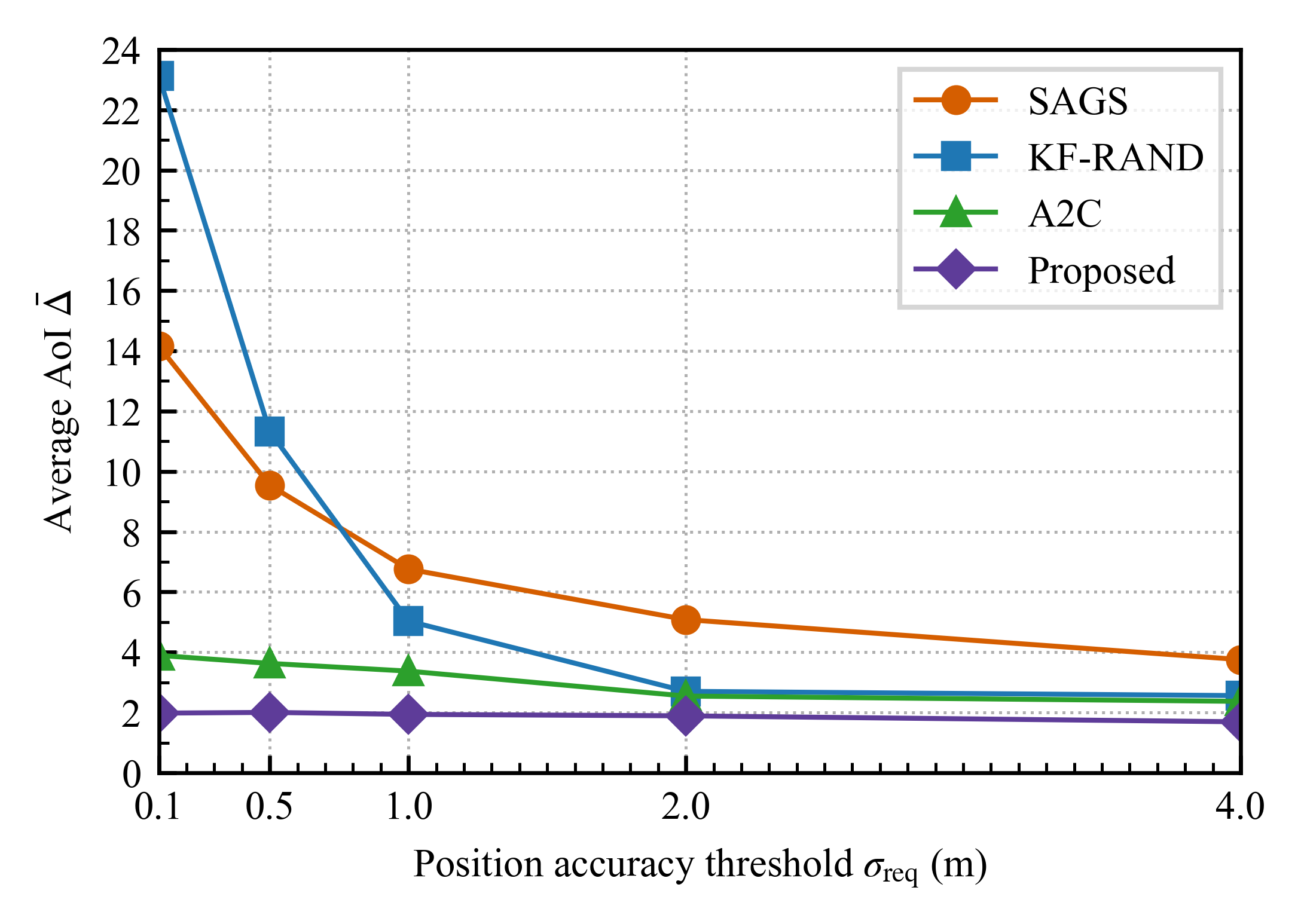}
        \caption{Average AoI versus position accuracy threshold (the maximum allowable 1-$\sigma$ horizontal position error) $\sigma_{\mathrm{req}}$.}
        \label{fig:aoi_vs_sigma}
    \end{figure}
    
    Fig.~\ref{fig:aoi_vs_sinr} shows the average AoI achieved by all schemes under different user SINR thresholds $\Gamma_{\mathrm{th}} \in \{0, 5, 10, 15, 20\}\,\mathrm{dB}$, with the UPA size fixed at $M_x \times M_y = 4 \times 4$ and the sensing accuracy requirement set to $\sigma_{\mathrm{req}} = 1\,\mathrm{m}$. As $\Gamma_{\mathrm{th}}$ increases, the increasingly stringent communication constraints result in deteriorated performance across all schemes. The AoI of SAGS remains nearly constant because it always serves the user with the largest AoI, but while neglecting the others, leading to poor overall freshness. In contrast, the proposed SAC-based method maintains the lowest AoI across all thresholds, demonstrating robust adaptability to increasingly stringent SINR constraints. A2C performs moderately well but consistently lags behind SAC, while KF-RAND achieves limited performance due to its lack of optimization. These results further validate the benefit of learning-based joint trajectory and beamforming control under varying link conditions.

    Fig.~\ref{fig:aoi_vs_sigma} presents the average AoI under varying position accuracy thresholds (the maximum allowable 1-$\sigma$ horizontal position error) $\sigma_{\mathrm{req}} \in \{0.1, 0.5, 1, 2, 4\}\,\mathrm{m}$, with six users, a $4 \times 4$ UPA, and $\Gamma_{\mathrm{th}} = 10\,\mathrm{dB}$ fixed. Smaller $\sigma_{\mathrm{req}}$ values impose stricter radar constraints, requiring the UAV to keep a close distance and allocate more power to the sensing target, which reduces the communication capacity and increases AoI. Despite the increasing sensing demand, the proposed SAC-based method maintains a consistently low AoI, demonstrating strong robustness and efficient resource allocation. A2C exhibits a similar downward trend, albeit with slightly elevated AoI levels compared to SAC. KF-RAND performs well when the requirement is loose but deteriorates rapidly as the constraint tightens. SAGS starts with high AoI under strict sensing and gradually improves as the radar burden becomes lighter, reflecting its limited adaptability.
    
    \begin{figure}[t]
        \centering
        \includegraphics[width=0.72\columnwidth]{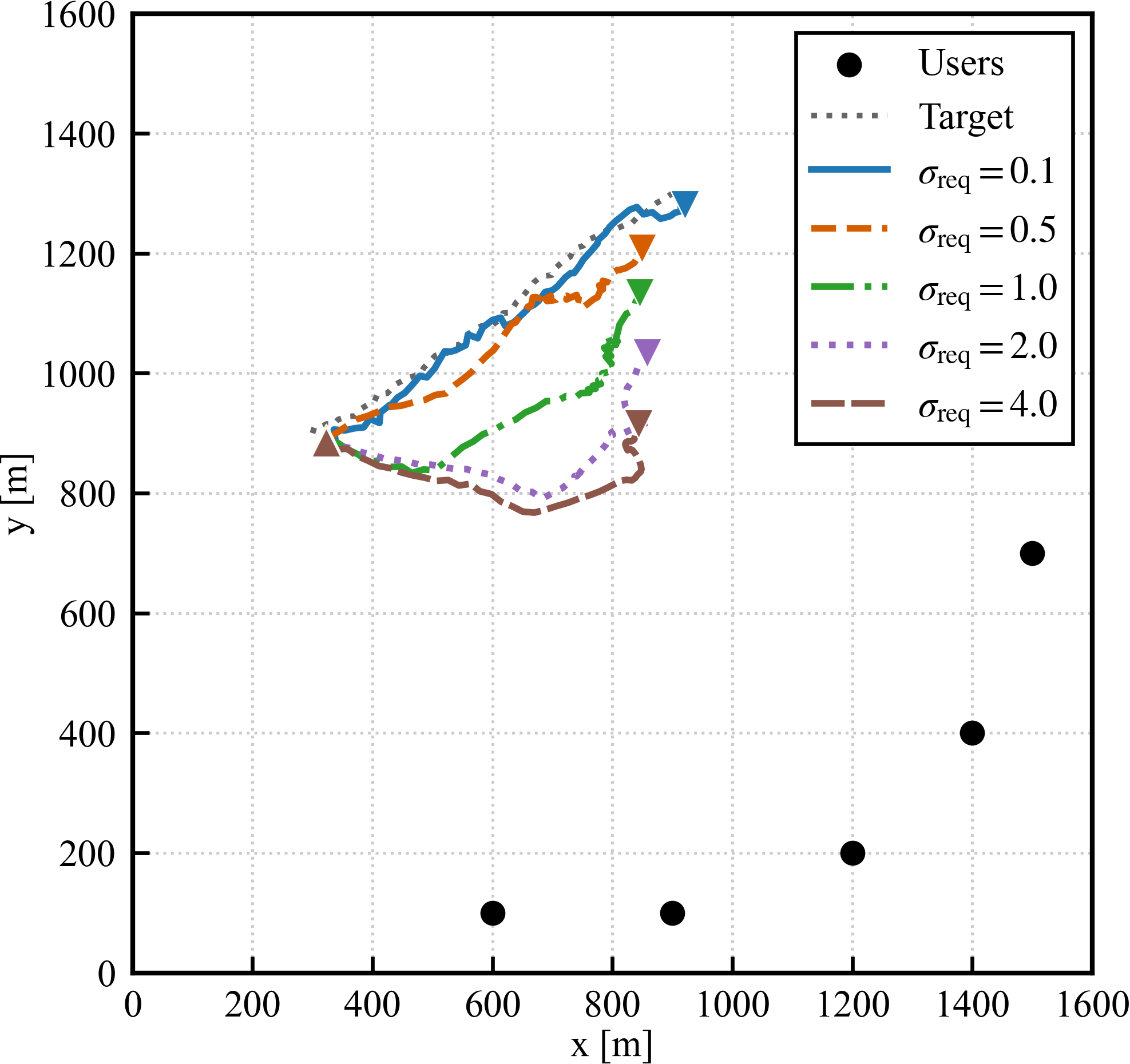}
        \caption{UAV trajectories under different position accuracy threshold $\sigma_{\mathrm{req}}$.}
        \label{fig:traj_sigma}
    \end{figure}
    
    \begin{figure}[t]
        \centering
        \includegraphics[width=0.87\columnwidth]{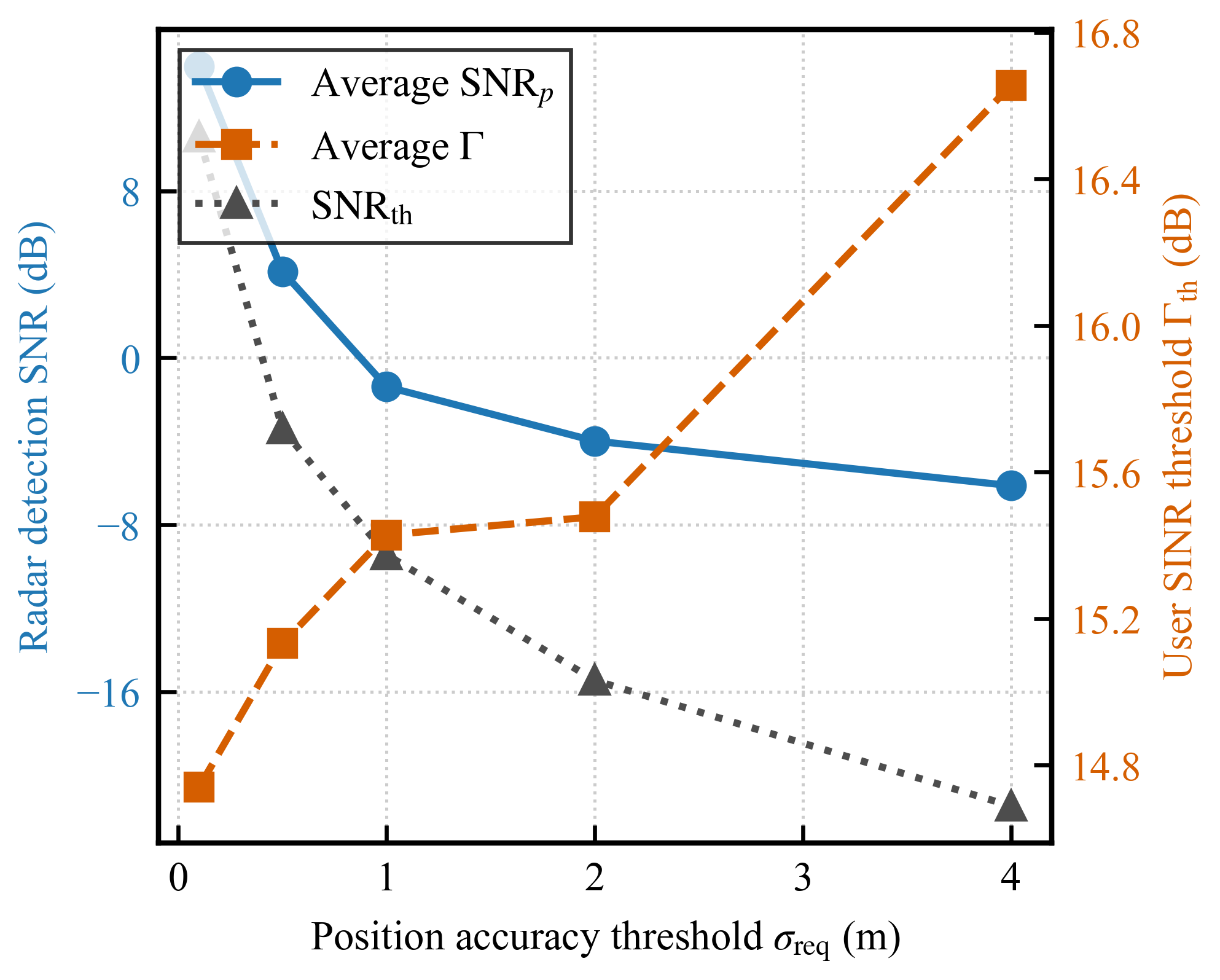}
        \caption{Average radar detection SNR $P_{r}$ and user SINR $\Gamma_{k}$ versus $\sigma_{\mathrm{req}}$.}
        \label{fig:snr_pr_sigma}
    \end{figure}
    
    \begin{figure*}[t]
    \centering
    \subfloat[]{\includegraphics[height=0.34\textwidth]{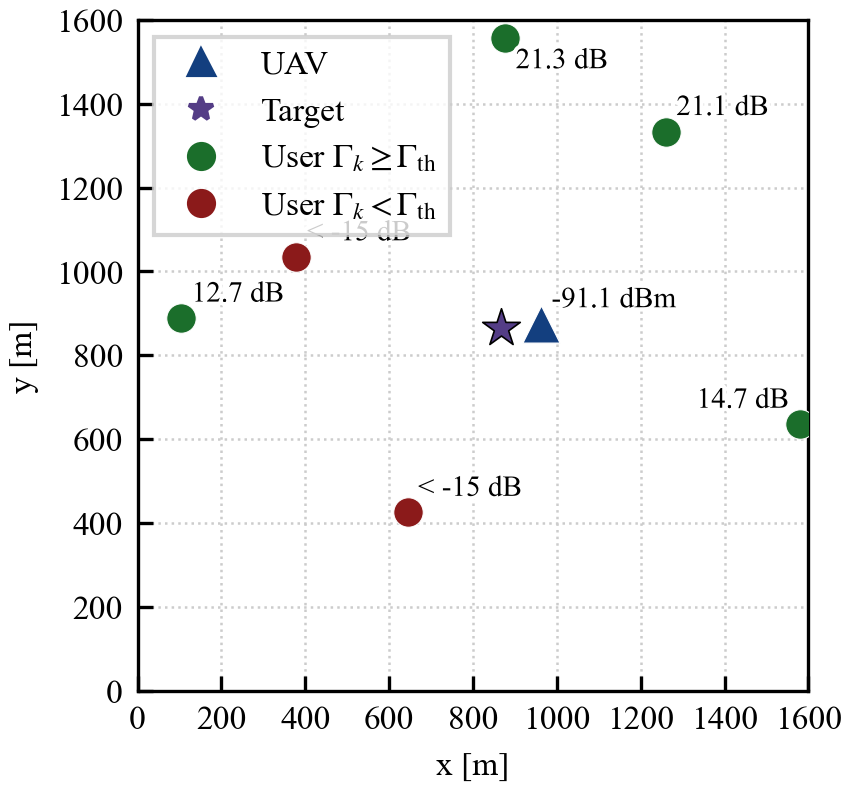}%
    \label{fig:uav:subfig:a}}
    \hfil
    \subfloat[]{\includegraphics[height=0.34\textwidth]{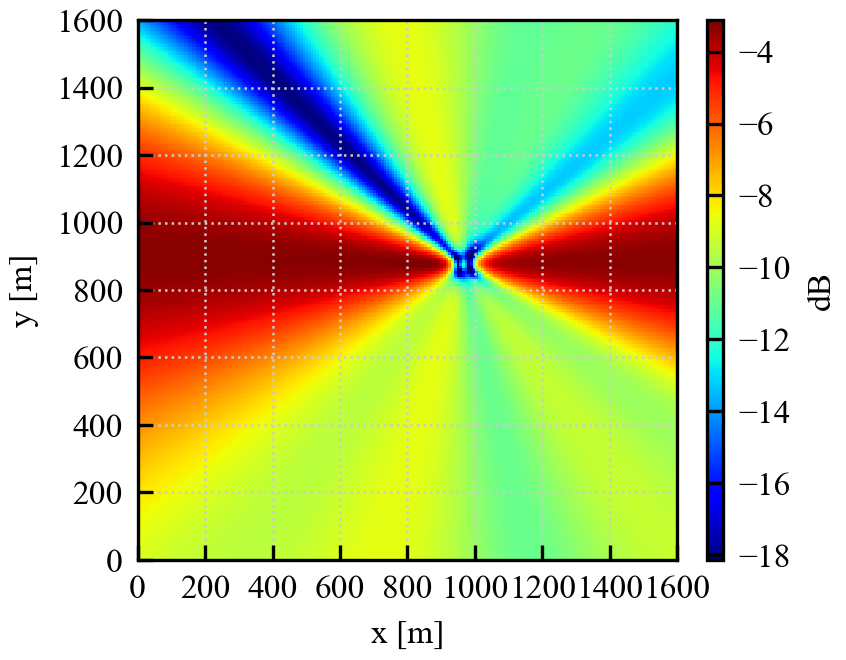}%
    \label{fig:uav:subfig:b}}
    \hfil
    \subfloat[]{\includegraphics[height=0.34\textwidth]{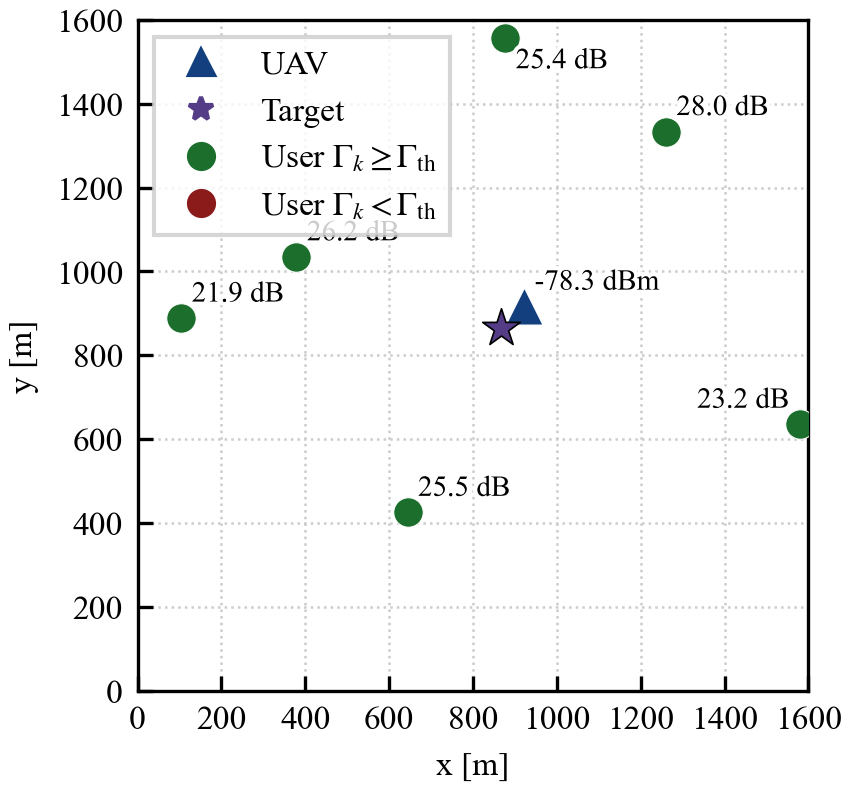}%
    \label{fig:uav:subfig:c}}
    \hfil
    \subfloat[]{\includegraphics[height=0.34\textwidth]{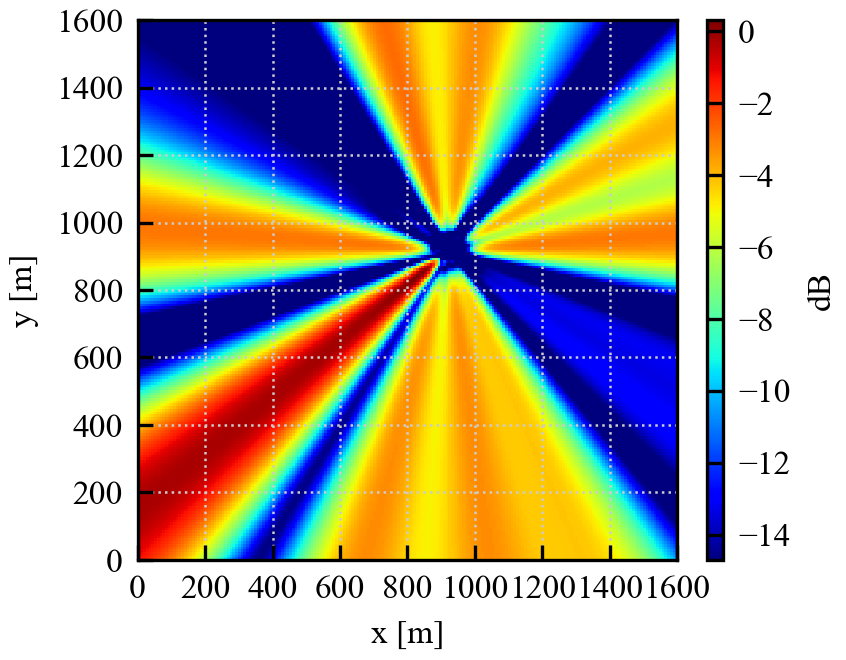}%
    \label{fig:uav:subfig:d}}
    \caption{Topology snapshots and ground-plane EIRP footprints at slot $n=40$ under two UPA sizes, $M_x=M_y=3$ and $M_x=M_y=6$. (a) Topology for $3\times3$ UPA; (b) EIRP footprint for $3\times3$ UPA; (c) Topology for $6\times6$ UPA; (d) EIRP footprint for $6\times6$ UPA.}
    \label{fig:uav_two_col}
    \end{figure*}

    To examine the effect of sensing position accuracy threshold $\sigma_{\mathrm{req}}$ on UAV behavior, we consider a scenario where six users are positioned far from the target's trajectory, specifically at coordinates [600, 100], [900, 100], [1,200, 200], [1,400, 400], [1,500, 700], and [1,500, 1,000] m. The target moves from [300, 900] m to [900, 1,300] m along an upward trajectory. This asymmetric layout emphasizes the trade-off between maintaining precise sensing and ensuring downlink communication. Fig.~\ref{fig:traj_sigma} depicts the SAC-based UAV flight trajectories obtained for five sensing position accuracy thresholds, i.e., $\sigma_{\mathrm{req}}\in\{0.1,0.5,1,2,4\}$ m, with six users, a $4 \times 4$ UPA, and $\Gamma_{\mathrm{th}} = 10\,\mathrm{dB}$ fixed. It is observed that a stringent requirement ($\sigma_{\mathrm{req}}\le0.5$ m) forces the UAV to track the target more closely, resulting in a narrow bundle of trajectories concentrated around the target trajectory. As $\sigma_{\mathrm{req}}$ relaxes, the trajectories progressively deviate downwards, forming a wider envelope. This deviation confirms the intuitive trade-off: when the localization constraint is loosened, the UAV shortens the average distance to the users located in the lower–right quadrant, thereby enhancing the communication link margin. The quantitative impact of~$\sigma_{\mathrm{req}}$ is summarized in Fig.~\ref{fig:snr_pr_sigma}. The radar‐receiving SNR, denoted by $\mathrm{SNR}_p$, and the user SINR, denoted by $\Gamma_{k}$, are first averaged in the linear domain over the $N$ slots and are then converted to decibels via $10\log_{10}(\cdot)$. On the left y-axis, the average radar detection $\mathrm{SNR}_p$ monotonically degrades from 13.97 dB to -6.12 dB as $\sigma_{\mathrm{req}}$ increases from 0.1 m to 4 m, while the required detection threshold $\mathrm{SNR}_{\mathrm{th}}$ remains lower than $\mathrm{SNR}_p$ over the entire range, ensuring reliable sensing. Conversely, the right y-axis shows that the average downlink users' SINR $\Gamma_{k}$ slightly rises from 14.74 dB to 16.65 dB. These two figures jointly confirm that loosening the sensing-accuracy constraint benefits downlink communication at the expense of radar performance, thereby visualizing the fundamental sensing-communication trade-off.

    To evaluate the effect of antenna array size on joint communication–sensing performance, Fig.~\ref{fig:uav_two_col} presents snapshots at slot $n=40$ from an identical simulation episode with fixed random seed. All system parameters are held constant—six static users, a sensing position accuracy threshold $\sigma_{\mathrm{req}} = 1$, a SINR threshold $\Gamma_{\mathrm{th}} = 10\,\mathrm{dB}$, except for the UPA size, which varies between $M_x=M_y=3$ and $M_x=M_y=6$. Subplots~(a) and~(c) illustrate the UAV, target, and user positions projected onto the horizontal plane. Users are colored green if their SINR $\Gamma_k$ exceeds the threshold, and red otherwise. The numeric labels near each user indicate the corresponding SINR $\Gamma_k$ (in dB), while the UAV label shows the received radar echo power $P_r$ (in dBm). Subplots~(b) and~(d) depict the equivalent isotropic radiated power (EIRP) footprints over the ground plane, expressed in dB, representing the post-beamforming transmit power distribution as if radiated isotropically. With $M_x=M_y=3$, the EIRP footprint in (b) shows a relatively broad mainlobe and higher sidelobes, reflecting the limited spatial resolution of the small array. This forces the UAV to compromise between user beams and the sensing beam, leading to only four users satisfying the SINR threshold in (a). In contrast, the $6\times6$ UPA in (d) achieves a narrower, higher-gain mainlobe and deeper nulls, enabling the SAC agent to simultaneously serve all users while maintaining a strong radar return, as observed in (c). This comparison highlights the substantial performance gains enabled by increasing the array size under the SAC policy.

    \begin{figure}[t]
        \centering
        \includegraphics[width=0.9\linewidth]{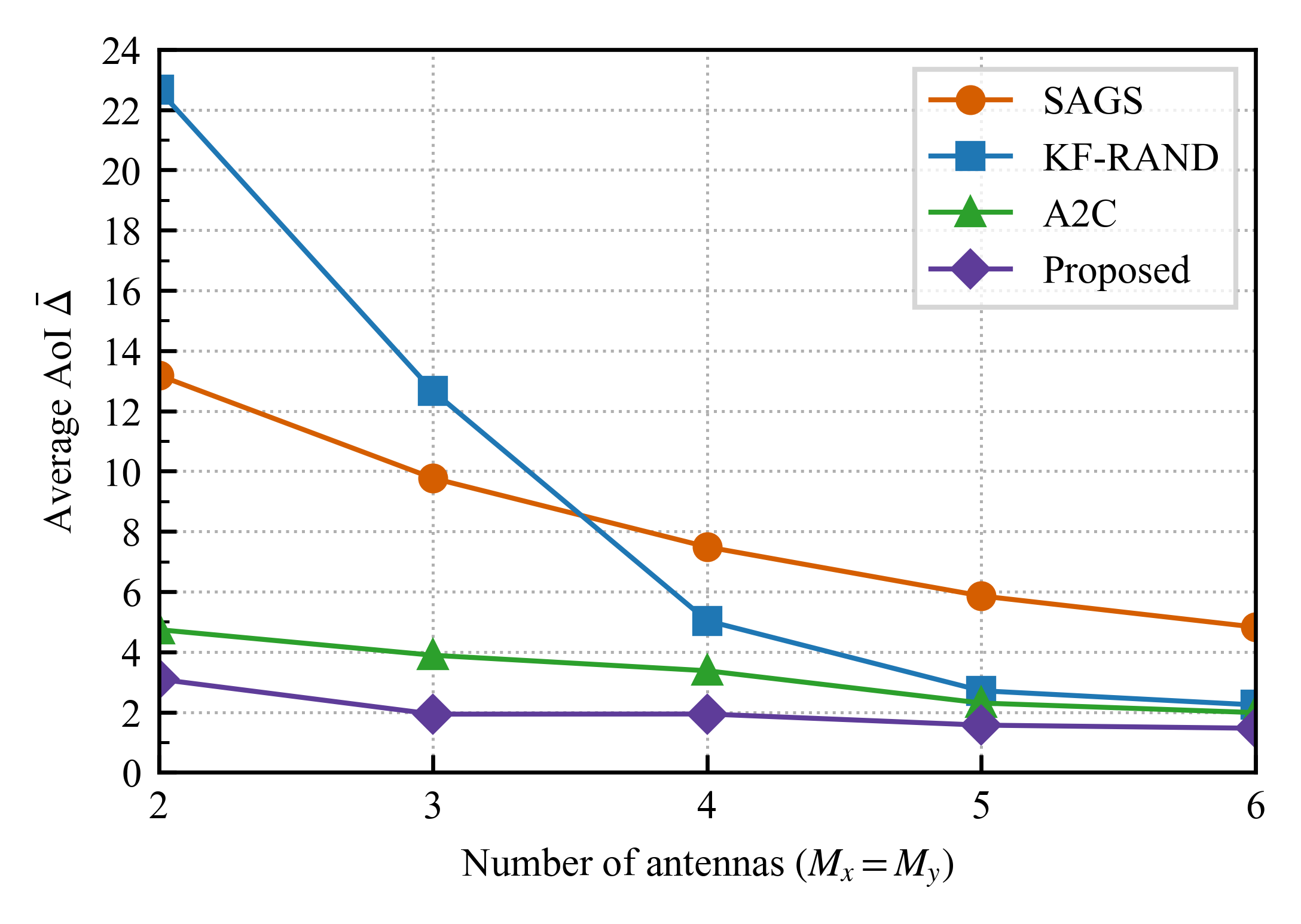}
        \caption{Average AoI versus UPA size $M_x=M_y$.}
        \label{fig:num_antennas}
    \end{figure}

    Fig.~\ref{fig:num_antennas} illustrates the average AoI as the UPA size \mbox{$M_x=M_y$} increases from 2 to 6, with other parameters held constant (position accuracy threshold $\sigma_{\mathrm{req}} =1 \mathrm{m}$, $4 \times 4$ UPA, and $\Gamma_{\mathrm{th}} = 10\,\mathrm{dB}$). The AoI consistently decreases across all evaluated schemes with the growth of array size, as a larger antenna array enhances both the communication and sensing capabilities. Notably, the proposed SAC method consistently achieves the lowest AoI due to its effective joint optimization of trajectory and beam control, fully exploiting the increased spatial degrees of freedom offered by larger arrays. Compared to KF-RAND, the advantage of SAGS is more pronounced for smaller arrays, where concentrating the available power towards fewer users significantly outperforms random allocation. However, this performance gap narrows as the array size increases, indicating that random allocation becomes increasingly effective with enhanced beamforming capabilities. Furthermore, A2C trails SAC across all array sizes, demonstrating that while it benefits from larger antenna arrays, it fails to match the effectiveness of SAC’s joint optimization strategy.

    \begin{figure}[t]
        \centering
        \includegraphics[width=0.9\linewidth]{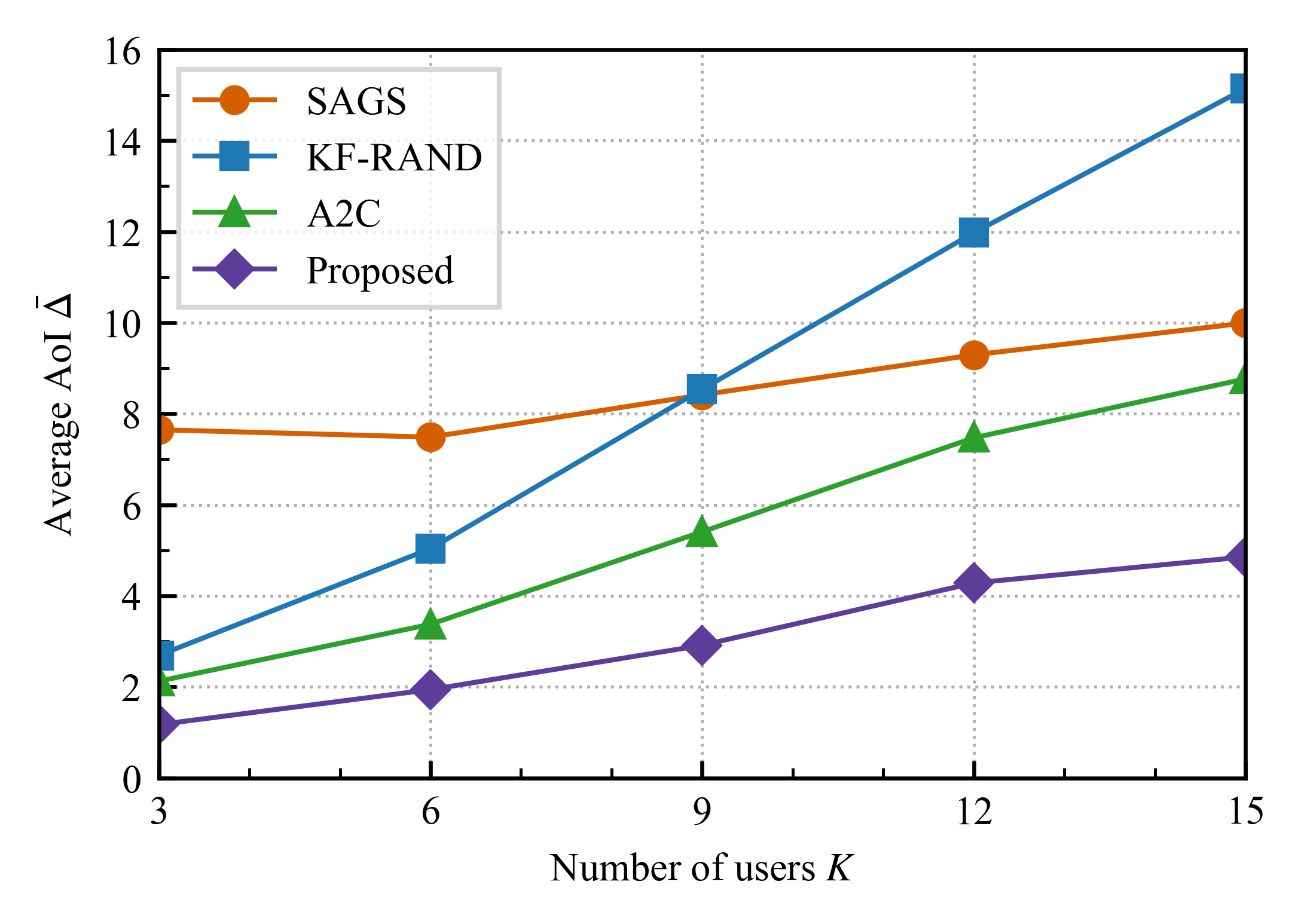}
        \caption{Average AoI versus number of users $K$.}
        \label{fig:num_users}
    \end{figure}
    
    Fig.~\ref{fig:num_users} illustrates the average AoI as the number of users $K$ increases from 3 to 15, while all other parameters remain fixed (the position accuracy threshold $\sigma_{\mathrm{req}} =1 \mathrm{m}$, $4 \times 4$ UPA, and $\Gamma_{\mathrm{th}} = 10\,\mathrm{dB}$). As expected, the average AoI grows monotonically with $K$ for each scheduling policy because more terminals contend for the same downlink resources. Across all values of $K$, the proposed SAC controller consistently achieves the lowest AoI, underscoring its ability to scale effectively through joint trajectory–beam optimization. In addition, SAGS outperforms KF-RAND at higher user loads ($K=12$ and $K=15$). Under these congested conditions, random power distribution in KF-RAND leaves many users below $\Gamma_{\mathrm{th}}$, whereas the single-user emphasis of SAGS guarantees at least one timely update per slot, thereby maintaining a lower overall AoI than purely stochastic allocation.

\section{Conclusion}
\label{sec:Conclusion}
This paper has presented a UAV-enabled ISAC system that leverages a superimposed transmit waveform for concurrent radar probing and downlink communication. This system has explicitly targeted an AoI-driven objective, focusing on tracking a moving target and the timely delivery of fresh information to multiple ground users. By integrating DRL with KF and RZF, the proposed solution has jointly optimized the UAV’s trajectory and multi-beam resource allocation. Extensive simulations have demonstrated that our SAC-based controller outperforms alternative methods in terms of average AoI across various system settings, including different user SINR thresholds, sensing position-accuracy requirements, antenna array sizes, and numbers of users. The results have highlighted critical performance trade-offs in UAV-ISAC, such as how tighter sensing requirements force the UAV to remain closer to the target at the expense of communication coverage, and how larger antenna arrays significantly improve beamforming resolution and overall AoI performance.

\small
\bibliographystyle{IEEEtran}
\bibliography{references}

\end{document}